\documentclass[acmsmall,nonacm]{acmart}
\makeatletter
\let\@authorsaddresses\@empty
\makeatother
\renewcommand\footnotetextcopyrightpermission[1]{}
\AtBeginDocument{%
  }

\usepackage{multirow}
\usepackage{color}
\usepackage{subfig}
\usepackage{bm}
\usepackage{hyperref}
\usepackage{cleveref}

\begin{document}

\title{WTDUN: Wavelet Tree-Structured Sampling and Deep Unfolding Network for Image Compressed Sensing}

\author{Kai Han}
\affiliation{
  \institution{Beijing Key Laboratory of Multimedia and Intelligent Software Technology, Beijing Institute of Artificial Intelligence, School of Information Science and Technology, Beijing University of Technology}
  \city{Beijing}
  \country{China}}

\author{Jin Wang}
\affiliation{
  \institution{School of Computer Science, Beijing University of Technology}
  \city{Beijing}
  \country{China}}

\author{Yunhui Shi}
\authornote{Yunhui Shi is the corresponding author.}
\affiliation{
  \institution{Beijing Key Laboratory of Multimedia and Intelligent Software Technology, Beijing Institute of Artificial Intelligence, School of Information Science and Technology, Beijing University of Technology}
  \city{Beijing}
  \country{China}}

\author{Hanqin Cai}
\affiliation{
  \institution{The Department of Statistics and Data Science and the Department of Computer Science, University of Central Florida}
  \city{Orlando}
  \country{USA}}

\author{Nam Ling}
\affiliation{
  \institution{The Department of Computer Science and Engineering, Santa Clara University}
  \city{Santa Clara}
  \country{USA}
}

\author{Baocai Yin}
\affiliation{
 \institution{Beijing Key Laboratory of Multimedia and Intelligent Software Technology, Beijing Institute of Artificial Intelligence, School of Information Science and Technology, Beijing University of Technology}
 \city{Beijing}
 \country{China}}

\renewcommand{\shortauthors}{Kai Han, Jin Wang, Yunhui Shi, Hanqin Cai, Nam Ling, and Baocai Yin}

\begin{abstract}
Deep unfolding networks have gained increasing attention in the field of compressed sensing (CS) owing to their theoretical interpretability and superior reconstruction performance. However, most existing deep unfolding methods often face the following issues: 1) they learn directly from single-channel images, leading to a simple feature representation that does not fully capture complex features; and 2) they treat various image components uniformly, ignoring the characteristics of different components. To address these issues, we propose a novel wavelet-domain deep unfolding framework named WTDUN, which operates directly on the multi-scale wavelet subbands. Our method utilizes the intrinsic sparsity and multi-scale structure of wavelet coefficients to achieve a tree-structured sampling and reconstruction, effectively capturing and highlighting the most important features within images. Specifically, the design of tree-structured reconstruction aims to capture the inter-dependencies among the multi-scale subbands, enabling the identification of both fine and coarse features, which can lead to a marked improvement in reconstruction quality. Furthermore, a wavelet domain adaptive sampling method is proposed to greatly improve the sampling capability, which is realized by assigning measurements to each wavelet subband based on its importance. Unlike pure deep learning methods that treat all components uniformly, our method introduces a targeted focus on important subbands, considering their energy and sparsity. This targeted strategy lets us capture key information more efficiently while discarding less important information, resulting in a more effective and detailed reconstruction. Extensive experimental results on various datasets validate the superior performance of our proposed method.

\end{abstract}


\maketitle

\section{Introduction}
Compressed sensing (CS) is a promising technique for signal acquisition and reconstruction \cite{ref01, ref02}. The target signal is first simultaneously sampled and compressed with linear random transformations. Then, the original signal can be reconstructed exactly from far fewer measurements than that required by the Nyquist sampling rate. Mathematically, a random linear measurement $\boldsymbol{y} {\in} {\mathbb{R}^M}$ can be formulated as $\boldsymbol{y}=\boldsymbol A\boldsymbol{x}$, where $\boldsymbol x {\in} {\mathbb{R}^N}$ is the original signal and $\boldsymbol A{\in}{\mathbb{R}^{M{\times}N}}$ is the sampling matrix with ${M {\ll}N}$. $r = M/N $ is the sampling rate (or CS ratio). To obtain a reliable reconstruction, traditional CS reconstruction methods commonly solve an optimization problem as follows:
\begin{equation}
\mathop{\mathbf{argmin}}\limits_{\boldsymbol{x}}{\;} \frac{1}{2}\Vert{\boldsymbol{y} - \boldsymbol A\boldsymbol{x}}\Vert_2^2 + \lambda\Vert{\Psi 
 \boldsymbol{x}}\Vert_1,
 \label{Ax-y}
\end{equation}
where $\Vert{\boldsymbol{y} - \boldsymbol A\boldsymbol{x}}\Vert_2^2$ denotes the data-fidelity term and $\Psi \boldsymbol{x}$ is the transform coefficients of $\boldsymbol x$  with respect to some transform $\Psi$, the sparsity of $\Psi \boldsymbol{x}$ is encouraged by the $\ell_1$-norm or other sparsity promoting norm with the regularization parameter $\lambda$. Currently, CS has been widely applied in various fields such as image reconstruction~\cite{ref43}, magnetic resonance imaging (MRI)~\cite{ref44}, snapshot-compressed imaging~\cite{ref45}, and communication.
\begin{figure*}[!t]
\centering
\includegraphics[width=5.0in]{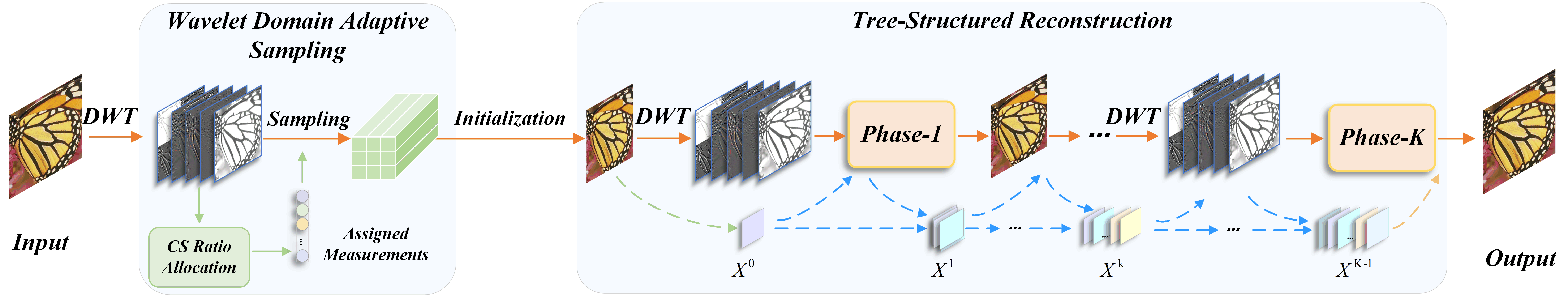}%
\caption{Illustration of WTDUN, which consists of a sampling module, an initial module, and $K$ reconstruction modules.}
\vspace{-0.50cm}
\label{frame}
\end{figure*}

Over the past decades, numerous efforts have been dedicated to image CS. There are two primary challenges in the field of CS: signal sampling and reconstruction. The construction of the sampling matrix $\boldsymbol A$ plays a crucial role in capturing the intrinsic structure of the original signal $\boldsymbol x$. Traditionally, $\boldsymbol A$ is often chosen as a random Gaussian, Poisson, or Toeplitz matrix to uniformly sample each component of the image. This can ensure accurate signal reconstruction with theoretical guarantees. However, the CS ratio is typically much lower than the Nyquist sampling rate, which usually results in artifacts. This phenomenon may be caused by the aliasing between high- and low-frequency information, which hinders detail and texture recovery during reconstruction. Therefore, it is important to treat different components distinctly to acquire more essential information during the sampling process.

Another key to achieving a superior reconstruction of $\boldsymbol x$ lies in the reconstruction algorithm. This has been the focus of numerous studies aimed at improving the quality of reconstructions in the past decades. Traditional optimization-based reconstruction methods often integrate image prior information~\cite{ref38, ref02-1, ref41,ref48}, such as sparsity in a specific transform domain, as a regularization term within the objective function. Consequently, these methods are capable of solving the sparse regularized optimization problem with precision. Although achieving satisfactory performance, these methods still have room for improvement due to their limited adaptability to diverse signals. Traditional CS methods usually suffer from challenges such as high computational complexity and the need for parameter tuning.

Recently, profiting from the powerful learning and fast computing abilities of deep networks, deep learning-based CS methods have garnered considerable interest. They can directly learn the inverse mapping from the CS measurement domain to the original signal domain~\cite{ref021, ref30}. Compared to traditional CS methods, deep learning-based CS methods dramatically reduce time complexity and greatly improve reconstruction performance. However, most existing deep learning-based methods are handcrafted designed, and trained as a black box, with limited insights in the CS domain. Thus, deep unfolding-based methods~\cite{ref04,ref25,ref3,ref36, ref6,ref35,ref53} are proposed, which are established by unfolding the first $K$ iterations of a referenced optimization algorithm and transforming all the steps of each iteration into learnable deep network components.
Deep unfolding methods can simultaneously maintain accuracy, interpretability, and speed by combining the advantages of traditional optimization-based algorithms and deep networks, such as ISTA-Net~\cite{ref04}, ADMM-CSNet~\cite{ref03} and MADUN~\cite{ref7}.

To address the above-mentioned issues, in this paper, we propose a novel wavelet domain-based deep unfolding framework. This framework introduces wavelet tree-structured reconstruction and wavelet-domain adaptive sampling, which can treat diverse components differently. By utilizing the rich information within the wavelet domain, our method is adapted to capturing intricate image features with greater efficiency. This ensures that the final image is not only visually pleasing but also fidelity to the structure of the original images. Specifically, we design a wavelet-domain adaptive sampling method that allocates CS measurements based on the difference of each subband, thereby customizing the sampling process to the varied features of the image. For CS reconstruction, we design a tree-structured prior to guiding our unfolding network, which can effectively exploit the inter-dependencies at different scales. It can capture key features across multiple scales, allowing the incorporation of information from one scale to enhance details at another scale. Since block partition breaks the global correlation of the whole image, a deblocking module is performed on the whole image in every reconstruction stage. This module serves to eliminate blocking artifacts and exploit contextual information between adjacent phases. Our extensive experimental results on diverse datasets demonstrate that our method can achieve better performance compared with other state-of-the-art CS methods. The main contributions of this paper are summarized as follows:

\begin{itemize}
\item  
A novel wavelet-domain deep unfolding framework for image compressed sensing is proposed, which achieves simultaneous wavelet domain adaptive sampling and tree-structured reconstruction by fully exploiting the structural sparsity of multi-scale wavelet coefficients.

\item  
A wavelet tree-structure prior guided unfolding network is designed, which effectively exploits the inter-dependencies among wavelet subbands at different scales. Through an iterative optimization process guided by wavelet tree-structured prior, accurate recovery of finer textures and sharper edges can be achieved. 
\item  
A wavelet domain adaptive sampling method that significantly improves sampling capabilities is developed. Unlike conventional methods that treat all subbands uniformly, our approach allows for a targeted focus on significant subbands, considering their energy and sparsity, which can capture relevant information effectively while discarding less important details. 

\item  
Extensive experimental results on various datasets validate the supreme performance of our proposed scheme.
\end{itemize}

\subsubsection*{Organization} The subsequent sections of this paper are organized as follows. Section 2 introduces related works of CS. Section 3 presents the network structure of our approach. Section 4 evaluates the performance of our methods and compares it to other state-of-the-art CS methods. At last, we conclude this paper in Section 5.

\section{Related Works}
The previous CS methods can be divided into two categories: traditional optimization-based methods and deep learning-based methods. In this section, we give a brief review.
\begin{figure*}[!t]
\centering
\includegraphics[width=5.3in]{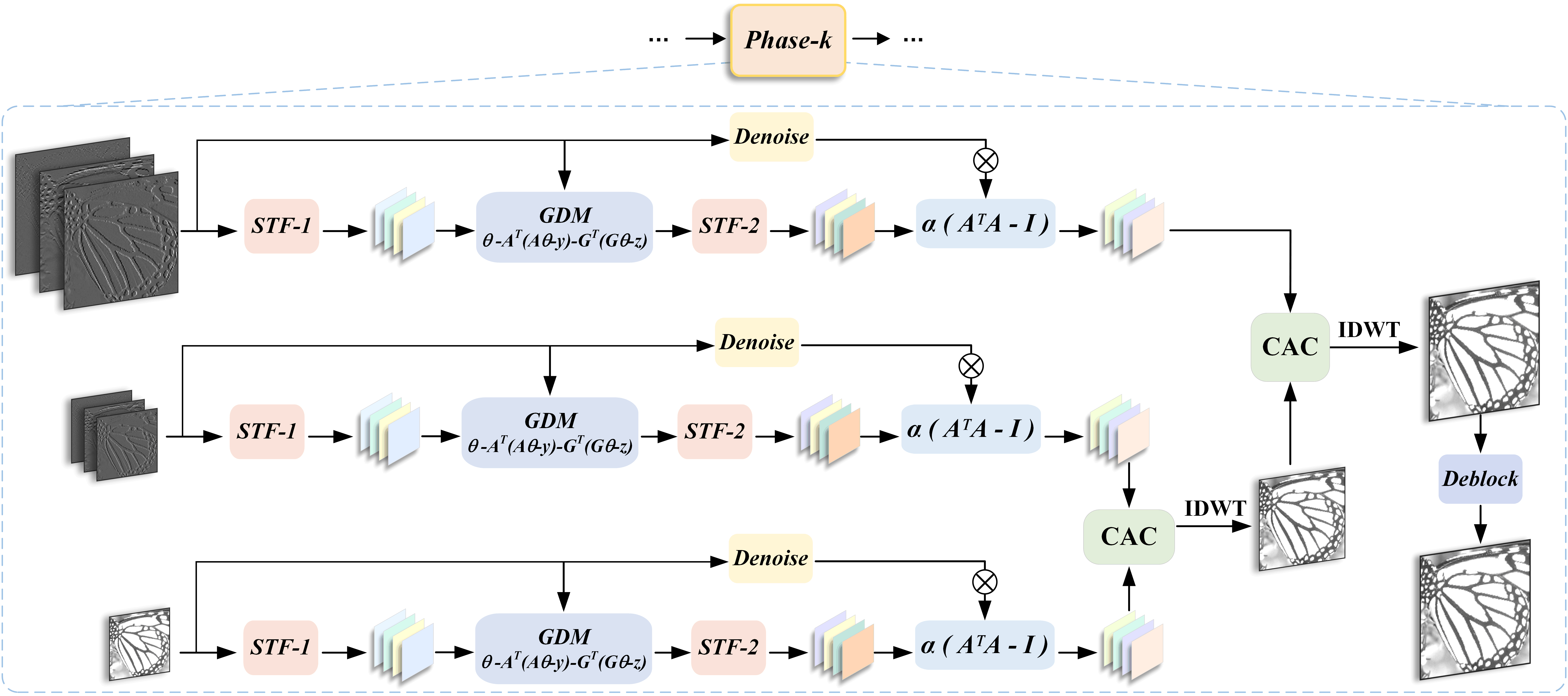}%
\caption{The detailed design of one single phase in WTDUN. The GDM(gradient descent module) represents Eq.~\eqref{gdm_r}. STF-1 and STF-2 denote Eq.~\eqref{relu_z} and \eqref{relu_theta} respectively. } 
\label{frame-stage}
\vspace{-0.65cm}
\end{figure*}
\subsection{Traditional Optimization-Based Methods}
\subsubsection{Sampling}
For sampling methods, they can be divided into \textit{uniform sampling} and \textit{adaptive sampling}. Uniform sampling methods typically use a random Gaussian, Poisson, or Toeplitz matrix to sample each image block and every component in blocks equally. On the contrary, adaptive sampling methods consider the differences in image structure. Recently, there has been extensive research on adaptive sampling methods, focusing on their capacity to enhance the overall performance of algorithms. Wang \textit{et al.}~\cite{ref02-2} use variance as a decision condition to split image blocks into two categories for sampling and reconstruction. Zhu \textit{et al.}~\cite{ref02-3} adaptively assign sampling measurements based on different statistical features in image blocks, which significantly improved the reconstruction quality. However, the above studies treat each component of image blocks equally, which may hinder the acquisition of high-quality reconstructed images.
\subsubsection{Reconstruction} 
In the field of CS reconstruction, most researchers have focused on two directions: image domain-based methods and wavelet domain-based methods. 

Methods categorized in the former group directly sample the original image blocks, which mainly include convex optimization algorithm~\cite{ref09}, greedy matching pursuit algorithm~\cite{ref010}, orthogonal matching pursuit algorithm~\cite{ref011}, Bayesian algorithm~\cite{ref012}, and gradient descent algorithm~\cite{ref013}. Afterward, to further improve the reconstruction performance, some elaborate priors have been applied to CS reconstruction, such as denoising prior, total variation prior, and group sparsity prior. Specifically, Li \textit{et al.}~\cite{ref02-1} develop a TV regularization constraint to improve the local smoothness. Zhang \textit{et al.}~\cite{ref017} present a group sparse representation to enhance image sparsity and exploit non-local self-similarity for image recovery. 

In the latter class of methods, the original image is first decomposed through a multi-layer wavelet transform. Subsequently, the resulting wavelet subbands are sampled in blocks. Finally, an inverse transform is applied to obtain the recovered image. The wavelet coefficients of images are organized in a quadtree structure. Most values of the wavelet coefficients cluster around zero, while only a small part shows significant amplitude. Such structure leads to strong sparsity which can greatly improve the imaging speed~\cite{ref08} and highly enhance reconstruction quality. For example, MS-BCS-SPL~\cite{ref015} is a CS algorithm to reconstruct multi-scale measurements in the wavelet domain, which can recover more details and sharper edges.  

Although traditional optimization methods have strong theoretical benefits, they still suffer from some issues such as excessive complexity, time-consuming, and poor real-time performance, which limit the broad application in practice.

\begin{figure*}
\centering
    \includegraphics[width=5.2in]{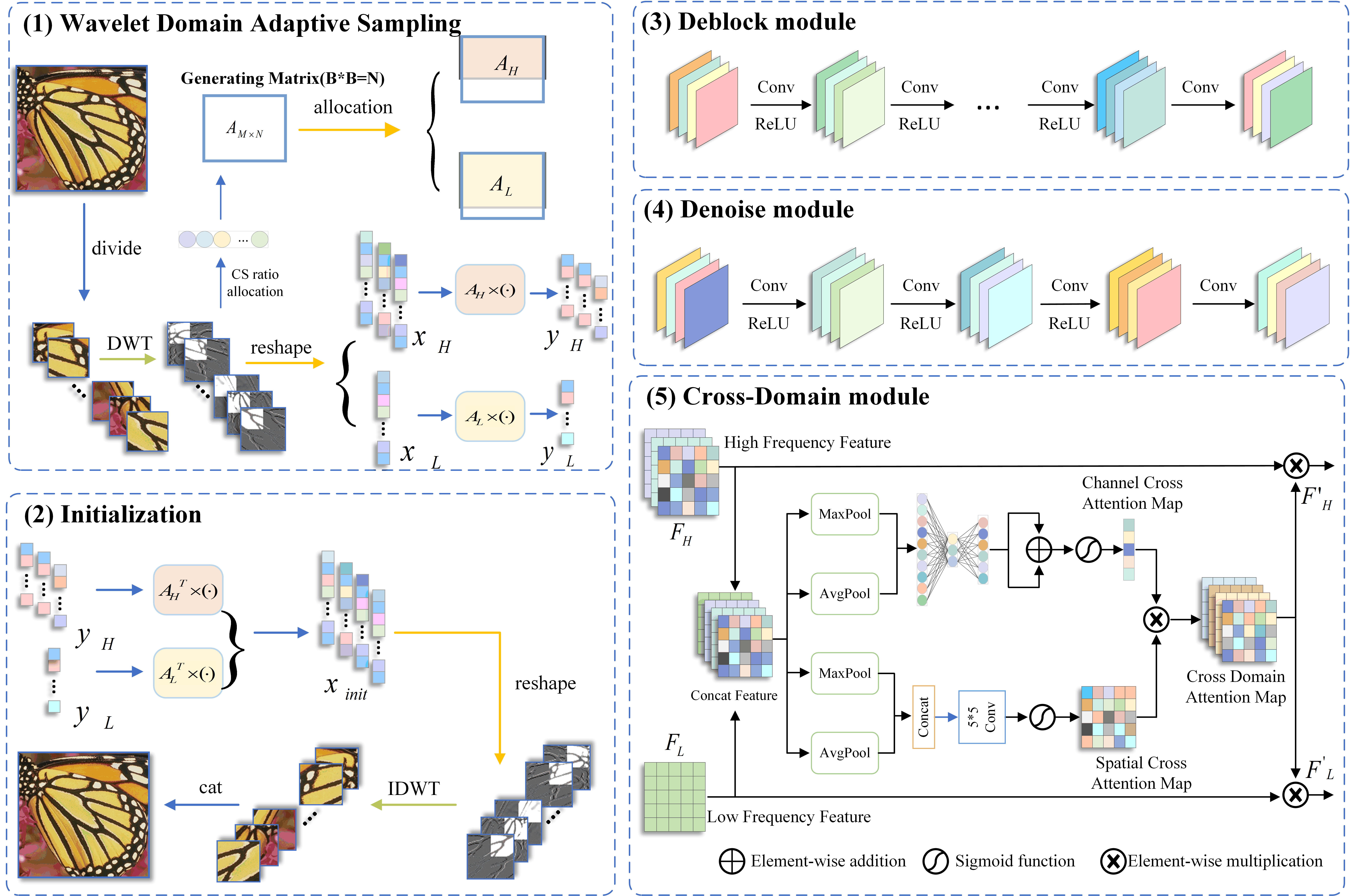}
    \caption{Detailed design of each module. (1) indicates the sampling process. (2) is the initial reconstruction process. (3) is the deblock module, which is composed of six convolution layers, with a ReLU activation layer between the adjacent convolution layers. (4) is the denoise module, which is composed of four convolution layers, with a ReLU activation layer between the adjacent convolution layers. (5) is the cross-subband fusion module (CAC).}
    \label{block}
    \vspace{-0.35cm}
\end{figure*}
\subsection{Deep Learning-Based Methods}
\subsubsection{Sampling} In contrast to traditional CS methods, some deep learning methods still use random matrix~\cite{ref3} to obtain CS measurements, while others adopt convolutional~\cite{ref021} or fully connected~\cite{ref020,ref27, ref28} layer to model sampling matrix. Recently, Zhou \textit{et al.}~\cite{ref4} propose a multi-channel model named BCS-Net using a channel-specific sampling network to realize adaptive CS ratio allocation. ~\cite{ref24} utilizes convolutional layers to extract high-dimensional features, which are subsequently compressed through a series of convolutional layers to acquire more sampling information. Although satisfactory reconstruction results can be achieved, there is still much room to improve the performance due to its lack of adaptability to the vast diversity of signals.

\subsubsection{Reconstruction}
As for deep learning-based reconstruction methods, they can be roughly categorized into deep network methods and deep unfolding methods. 

\textbf{Deep Network Methods:} 
For classical deep network methods, Mousavi \textit{et al.}~\cite{ref019} first exploit a stack denoising autoencoder (SDA) to recover images from CS measurements. Afterward, Lohit \textit{et al.}~\cite{ref020} design a CNN-based model called ReconNet to improve the image reconstruction quality. Shi \textit{et al.}~\cite{ref021} propose a residual CNN-based framework dubbed CSNet to jointly train the sampling matrix and the reconstruction network. Huang \textit{et al.}~\cite{ref31,ref32} design a reconstruction network to acquire high-quality reconstruction by exploiting the Gaussian Scale Mixture prior. \cite{ref022} uses generative adversarial networks (GAN) for CS reconstruction. \cite{ref49} successfully improves the reconstruction quality through the analysis of image features in the frequency domain. DDS-Net~\cite{ref52}, a novel dynamic network, efficiently customizes denoising processes for various image noise levels. Moreover, the application of transformer in image reconstruction has also yielded significant achievements~\cite{ref15,ref47}.

\textbf{Deep Unfolding Methods:} 
Deep unfolding methods realize mathematical optimization iterations by designing deep networks. Therefore, deep unfolding methods combine the advantages of traditional iterative optimization methods and classical deep network methods. For example, Zhang \textit{et al.}~\cite{ref04} extend the popular algorithm ISTA to ISTA-Net for optimizing a $\ell_1$-norm CS model. Afterwards, Zhang \textit{et al.}~\cite{ref3} propose a CS model called AMP-Net by unfolding the denoising process of the approximate message-passing algorithm, incorporating a deblocking module in each stage to remove blocking artifacts. Metzler \textit{et al.}~\cite{ref023} design LDIT and LDAMP from the denoise-based iterative thresholding algorithm. Yang \textit{et al.}~\cite{ref03} propose a deep learning framework named ADMM-CSNet by the redesign of CS magnetic resonance imaging (CS-MRI). 

Compared with traditional iterative optimization-based methods, deep learning-based methods greatly reduce computational complexity and time complexity. Although the above deep learning-based methods have achieved excellent success in the field of CS, there are still some issues that texture details are not recovered well enough and no sparsity prior knowledge is combined.

To address the above-mentioned issues, we propose a novel wavelet domain-based deep unfolding framework to simultaneously enable wavelet domain adaptive sampling and tree-structured reconstruction. In the sampling process, we present a wavelet domain adaptive sampling method to allocate CS measurements based on the different importance of multi-scale subbands. The reconstruction process adopts a tree-structured prior guided model that takes advantage of the structure sparsity in multi-scale subbands. This model enables the recovery of more structural and finer detail components.

\section{Proposed Method}  
\subsection{Overall Architecture}
 As shown in Fig.~\ref{frame}, WTDUN is an end-to-end deep unfolding model that integrates an optimization algorithm with a deep network, and it comprises a sampling module, an initialization module, and $K$ reconstruction stages. Traditional optimization algorithm \cite{ref33,ref34} solves the CS reconstruction problem by iterating between the following two update steps:
\begin{align}
\label{amp_1}
   \boldsymbol{z}^{k-1} &= \boldsymbol{y} - \boldsymbol A\boldsymbol{x}^{k-1}, \\
\label{amp_2}
   \boldsymbol{x}^{k} &= \zeta_k(\boldsymbol{A}^T\boldsymbol{z}^{k-1} + \boldsymbol{x}^{k-1} ),
\end{align}
 where $k$ is the number of iterations, $\boldsymbol{A}^T$ is the transpose of sampling matrix $\boldsymbol A$, and $\zeta(\cdot)$ is an ideal image denoiser with a denoising perspective~\cite{ref3}: 
\begin{equation}
\label{denoise}
    \boldsymbol{A}^T\boldsymbol{z}^{k-1} + \boldsymbol{x}^{k-1} = \overline{\boldsymbol{x}} + \boldsymbol{e},
\end{equation}
where $\boldsymbol{e} = (\boldsymbol{A}^T\boldsymbol{A}-\boldsymbol{I})(\overline{\boldsymbol{x}}-\boldsymbol{x}^{k-1})$ denotes a noise term. The clean signal can be obtained by calculating:
\begin{equation}
\label{amp_deniose}
    \boldsymbol{x}^k = \boldsymbol{A}^T\boldsymbol{z}^{k-1} + \boldsymbol{x}^{k-1} - (\boldsymbol{A}^T\boldsymbol{A}-\boldsymbol{I})(\overline{\boldsymbol{x}}-\boldsymbol{x}^{k-1}).
\end{equation}
By combining with the wavelet tree-structure prior, we transform Eq.~\eqref{amp_deniose} into the $k$-th reconstruction stage of our network, as shown in Fig.~\ref{frame-stage}. Considering the different importance of image high- and low-frequency components, a wavelet domain adaptive sampling module is constructed to enable adaptive sampling as in Fig.~\ref{block}. Then in subsections B, C, and D, each module will be elaborated separately.

\subsection{Wavelet Domain Adaptive Sampling}
\subsubsection{CS measurement Ratio Allocation}
The CS measurement ratio is usually significantly lower than the Nyquist sampling rate, leading to undesirable artifacts. This phenomenon can be attributed to aliasing between high- and low-frequency information, which may prevent the accurate recovery of image details and textures during reconstruction. To mitigate the problem, we employ a Haar wavelet to separate the image into different frequency components. Subsequently, CS measurements for each component are assigned based on their importance.
\begin{equation}
\label{dwt_x}
{\boldsymbol \theta_s} = \Psi \boldsymbol{X},
\end{equation}
where $\Psi = [\boldsymbol{\psi}_1, \boldsymbol{\psi}_2,\cdots, \boldsymbol{\psi}_n]$ is the transform basis, $s\in\{LL_l,HL_i,LH_i,HH_i\}$, $1\leq i\leq l$, $i$ means the wavelet decomposition level, $\boldsymbol \theta$ means each wavelet subband of image $\boldsymbol X$.
After multi-level wavelet decomposition, we obtain a sequence of subbands. Based on the property that the lowest frequency in the wavelet domain can carry the image energy and the high-frequency subbands can express the image textures and details, we design an algorithm to realize adaptive CS ratio allocation. We use the mean to evaluate subband energy and the standard deviation to measure subband sparsity.

In the allocation algorithm, we initially calculate the standard deviation $\sigma$ and mean $\mu$ of absolute values $ |\boldsymbol \theta| $ for each wavelet subband. Subsequently, we introduce $\eta$ as a control parameter to adjust the contribution of $\sigma$ and $\mu$ when determining the allocation weight. The allocation weight $W$ is obtained by summing up the products of $\sigma$, $\mu$, and their respective control parameters. Next, we compute the percentages $P_H$ and $P_{LL}$ representing high-frequency ($W_H$) and low-frequency ($W_{LL}$) contributions, respectively.
By utilizing the percentages $P_H$ and $P_{LL}$, we can derive the corresponding quantities of CS measurements. Subsequently, an upper bound $\Theta$ is established to prevent sample overflow. In cases where the number of low-frequency measurements surpasses this threshold $\Theta$, it is adjusted to match the number of low-frequency CS measurements. Any additional CS measurements are then allocated to other wavelet subbands. Given that high-frequency subbands in images exhibit a tree structure following multi-level wavelet decomposition, these subbands at different scales demonstrate similar structures. Rather than considering sparsity when assigning measurements to high frequencies, our focus lies solely on their energy levels. Consequently, CS measurements are allocated among them based on their respective means at different levels. Finally, we set a bias term ($ \sum\limits_{i=0}^n {bias_i} = 0 $) to increase the flexibility of this allocation method.

\subsubsection{Block-by-block Sampling}
 We employ the block-by-block sampling method to independently sample each non-overlapping image block with a fixed size of $n \times n$. Initially, an image $\boldsymbol{X} \in \mathbb{R}^{H\times W}$ is divided into blocks $\boldsymbol{x}_j \in \mathbb{R}^{n \times n}, j \in \{1,2,\cdots ,J\}$. Subsequently, these blocks are decomposed into multi-scale wavelet subbands $\boldsymbol{\theta}_{s} \in \mathbb{R}^{n_s\times n_s}$. Before sampling, each subband is vectorized as a vector. As illustrated in Fig.~\ref{block}, the sampling process of each subband $\boldsymbol{\theta}_{s}$ at decomposition level $i$ within block $\boldsymbol{x}_{j}$ can be described as follows:
\begin{equation}
\label{y_sample}
\boldsymbol y_{s} = \boldsymbol{\Phi_{s}\Psi}\boldsymbol{x}_{j} = \boldsymbol{A}_{s}\boldsymbol{\theta}_{s}, 
\end{equation}
where $ \boldsymbol{y}_{s}\in\mathbb{R}^{M_s \times J} $ is the CS measurements of $\boldsymbol{x}_j$ and $H\cdot W = J \cdot n^2$, $s\in\{LL_l,HL_i,LH_i,HH_i\}$, $1\leq i\leq l$, $i$ means the wavelet decomposition level. $\boldsymbol{A}_{s}\in\mathbb{R}^{M_s\times n^2}$ is the multi-scale sampling operator, which consists of two components (a multi-scale transform $\boldsymbol{\Psi}$ and a multi-scale block-based measurement process $\boldsymbol{\phi}$ ). $\boldsymbol{\Psi}$ produces $\textit{L}$ levels after multi-level wavelet decomposition, therefore, $\boldsymbol{\phi}$ consists of $\textit{L}$ different block-based sampling operators, one for each level.

\subsection{Wavelet Tree-Structure Prior Guided Reconstruction}
The reconstruction module exploits the quad-tree representation of wavelet coefficients to capture the inter-dependencies among coefficients at different scales in an image. This representation enables us to identify groups of coefficients that can either be zero or nonzero simultaneously, thereby providing valuable structural insights for enhancing image reconstruction quality. Specifically, the wavelet tree-structured prior guided reconstruction model can be described as follows:

\begin{equation}
\label{tree}
\mathop{\mathbf{argmin}}\limits_{\boldsymbol \theta}{\;} \frac{1}{2}\Vert{\boldsymbol{y} - \boldsymbol{A\theta}}\Vert_2^2 + \beta{(\Vert \boldsymbol \theta\Vert_1 + \Vert \boldsymbol{G\theta}\Vert_2)},
\end{equation}
where $\boldsymbol G$ represents the set of all parent-child groups for the wavelet tree. And we adopt weighted $\ell_{1}$- and $\ell_2$-norms in the second term in Eq.~\eqref{tree} to impose the tree-structured sparsity within wavelet subbands.
\subsubsection{Initial Reconstruction Module}
The initial reconstruction module is illustrated by Eq.~\eqref{x_init}. The wavelet coefficients of the original image can be initially reconstructed from its multi-scale measurements as shown in Fig.~\ref{block}:
\begin{equation}
\label{x_init}
\boldsymbol \theta_{s} = {\boldsymbol A_{s}}^T \boldsymbol y_{s},
\end{equation}
where $s\in\{LL_l,HL_i,LH_i,HH_i\}$, $1<i<L$. There is only one low-frequency subband, so $LL$ is not annotated with subscripts " $i$ ". ${\boldsymbol A_{s}}^T \in \mathbb{R}^{n^2\times M_s}$ represents the linear mapping matrix and it is the pseudo-inverse matrix of $\boldsymbol A_s$. Then we transform $\boldsymbol \theta_s$ into full-image $\boldsymbol X^0$ by inverse wavelet transform.

\subsubsection{Deep Reconstruction Module}
The deep reconstruction process is performed on the initial reconstruction result $\boldsymbol X^0$ and improves its quality. We divide the reconstruction model into $K$ stages. Each stage alternatively implements the projection in the wavelet domain and the full-image deblocking in the spatial domain. With $\boldsymbol{z} = \boldsymbol{G\theta}$ constraint, Eq.~\eqref{tree} can be rewritten as:
\begin{equation}
\label{min_tree}
\mathop{\mathbf{argmin}}\limits_{\boldsymbol \theta}{\;} \frac{1}{2}\Vert{\boldsymbol y - \boldsymbol{A\theta}}\Vert_2^2 + \beta{(\Vert \boldsymbol \theta\Vert_1 + \Vert \boldsymbol z\Vert_2) + \frac{\lambda}{2}\Vert \boldsymbol z - \boldsymbol{G\theta}\Vert_2^2}.
\end{equation}
The variables $\boldsymbol z$ and $\boldsymbol \theta$ are coupled together in the minimization of Eq.~\eqref{min_tree}, thus solving them simultaneously is computationally intractable. Therefore, we divide the problem into two sub-problems, namely $\boldsymbol \theta$-subproblem and $\boldsymbol z$-subproblem, which are described as follows.

\begin{figure*}[!t]
\centering
\captionsetup[subfloat]{labelsep=none,format=plain,labelformat=empty}
GT \ \quad COAST\cite{ref5}  \ ISTA-Net++\cite{ref6}  \ OPINE\cite{ref2}   \ DPUNet\cite{ref11} \ AMP-Net\cite{ref3} \quad Ours  \qquad \
\\
\subfloat[PSNR(dB)/SSIM]{\includegraphics[width=0.78in]{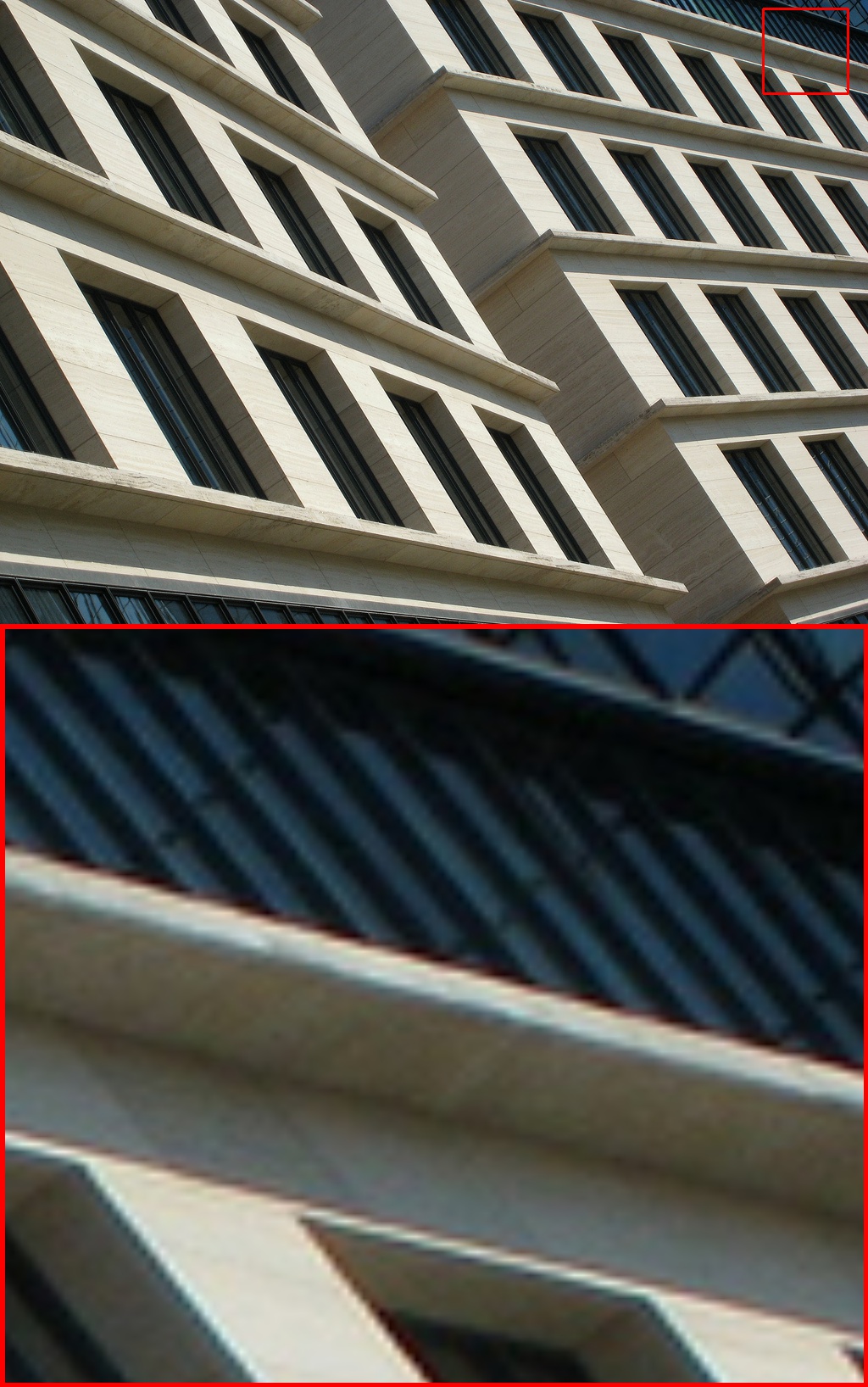}
}
\hfil
\subfloat[{\textit{33.53}}/0.8972]{\includegraphics[width=0.78in]{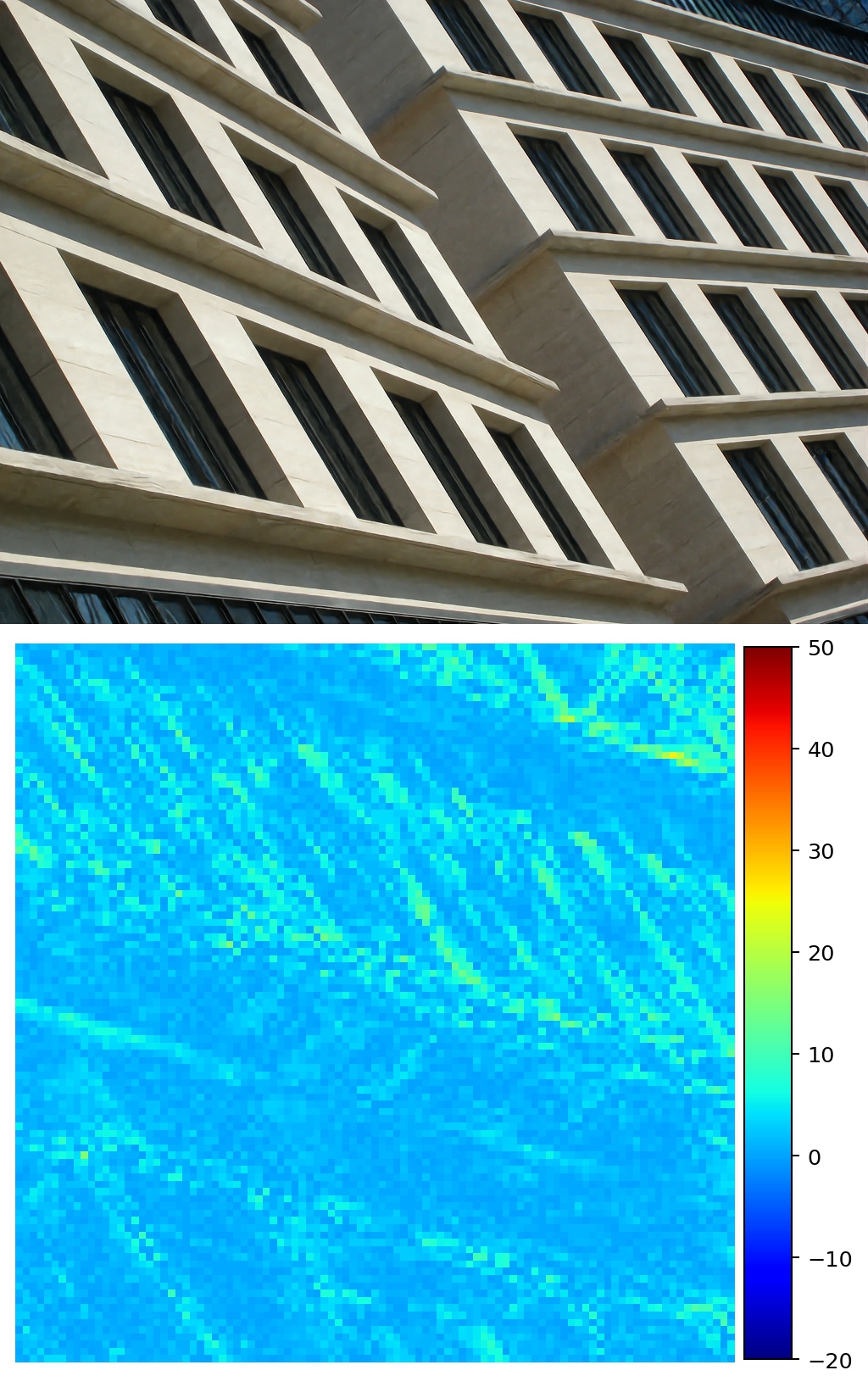}
}
\hfil
\subfloat[31.46/0.8602]{\includegraphics[width=0.78in]{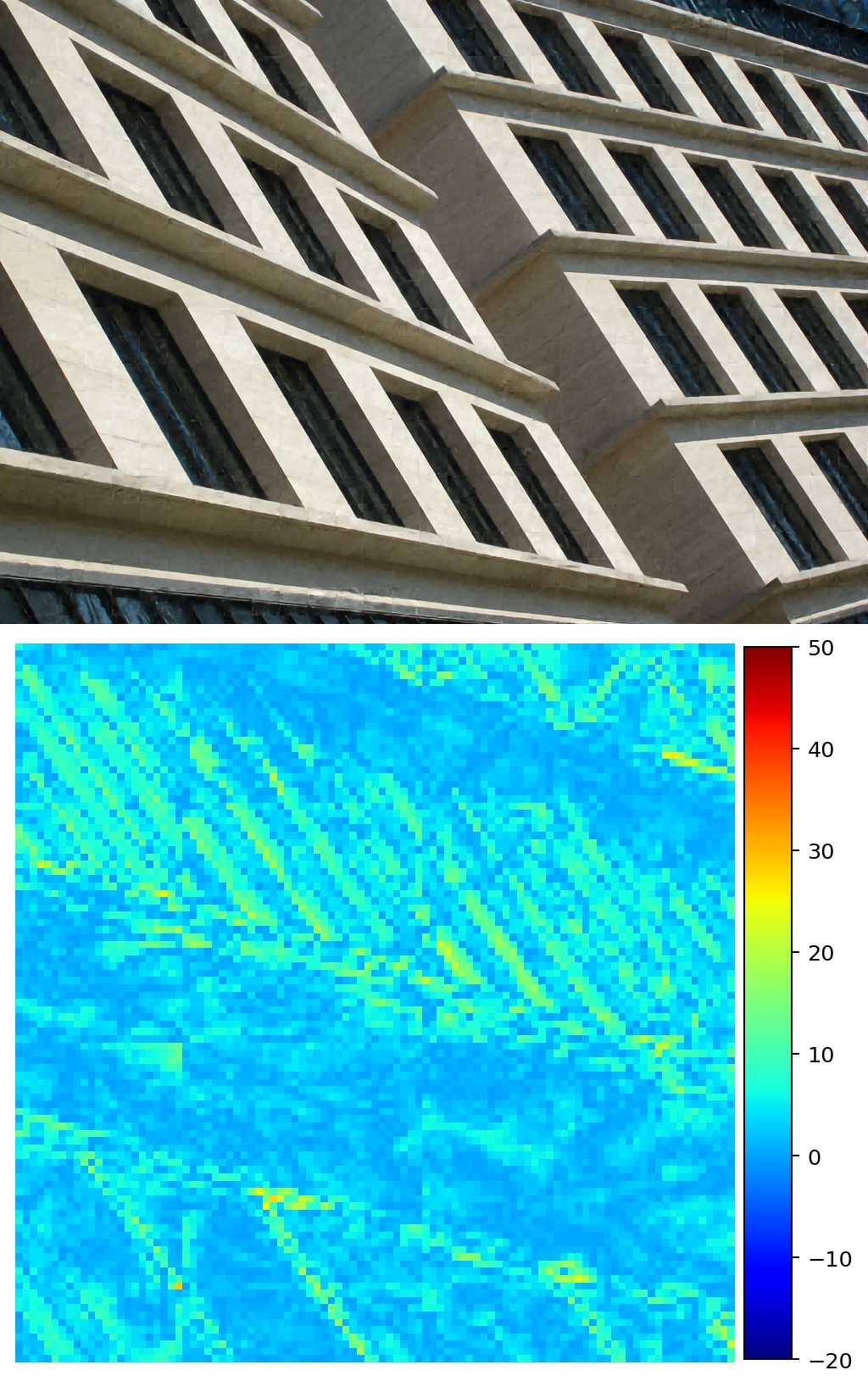}
}
\hfil
\subfloat[32.79/\textit{0.9059}]{\includegraphics[width=0.78in]{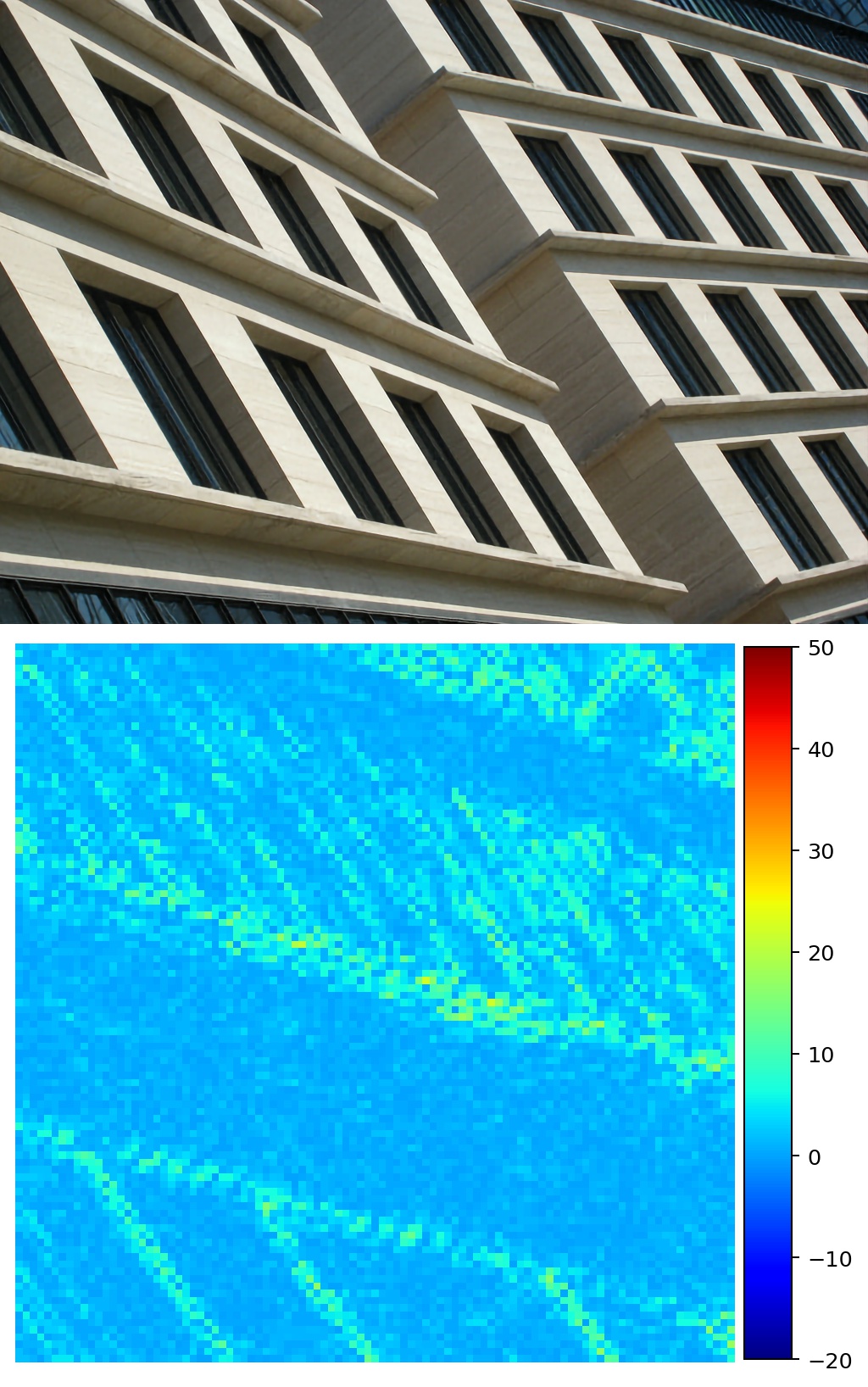}
}
\hfil
\subfloat[31.86/0.8960]{\includegraphics[width=0.78in]{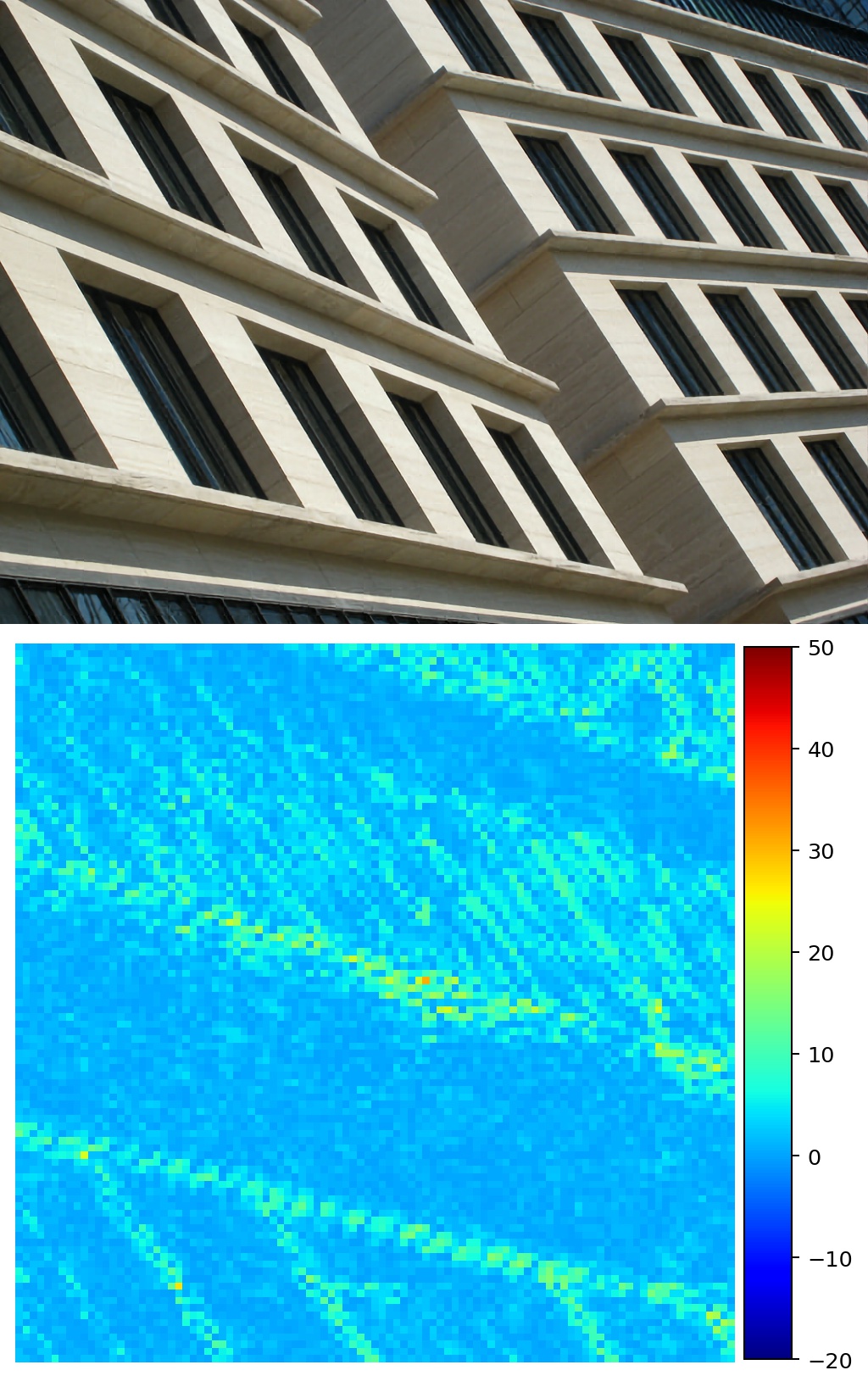}
}
\hfil
\subfloat[31.40/0.8908]{\includegraphics[width=0.78in]{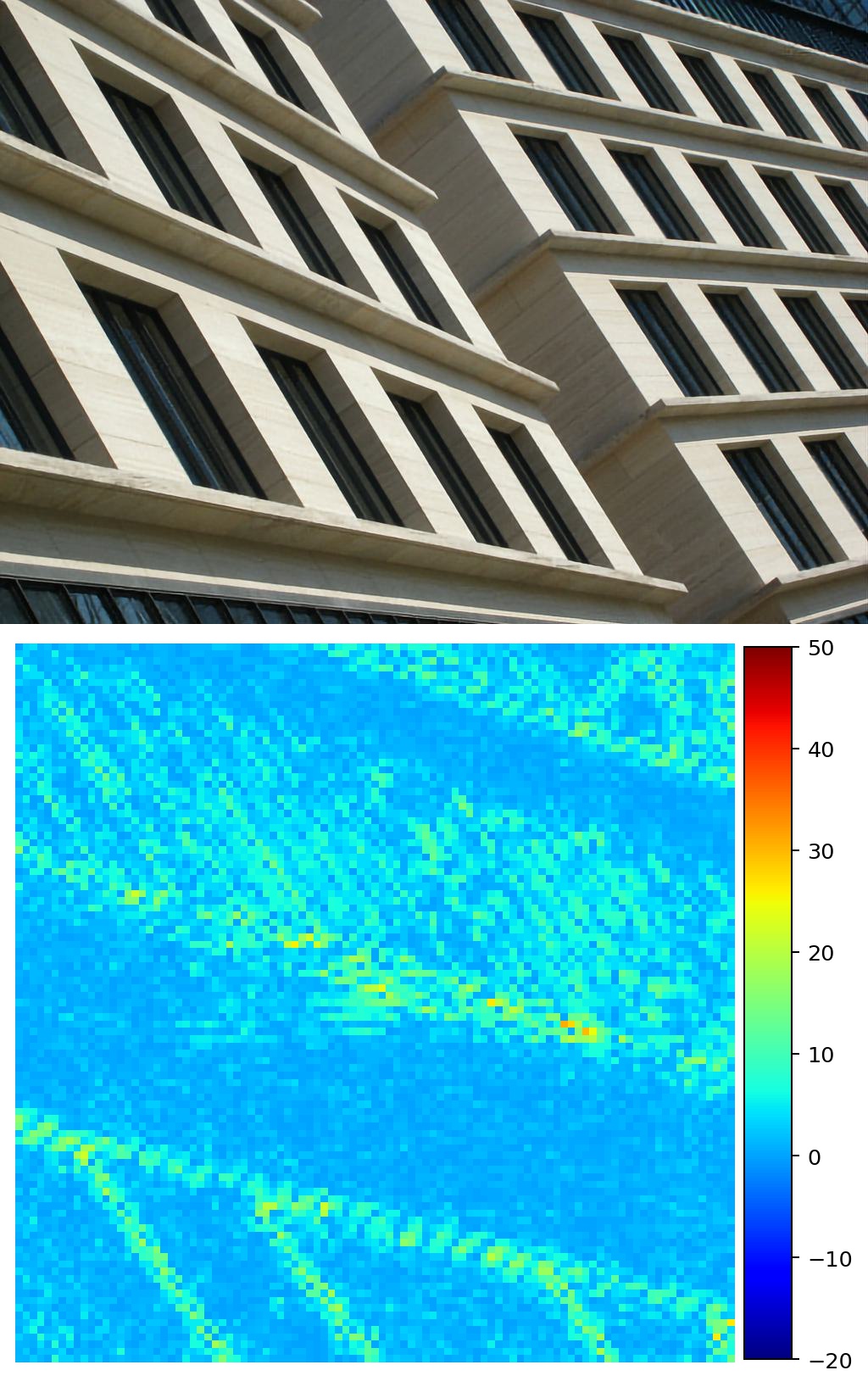}
}
\hfil
\subfloat[\textbf{33.79}/\textbf{0.9099}]{\includegraphics[width=0.78in]{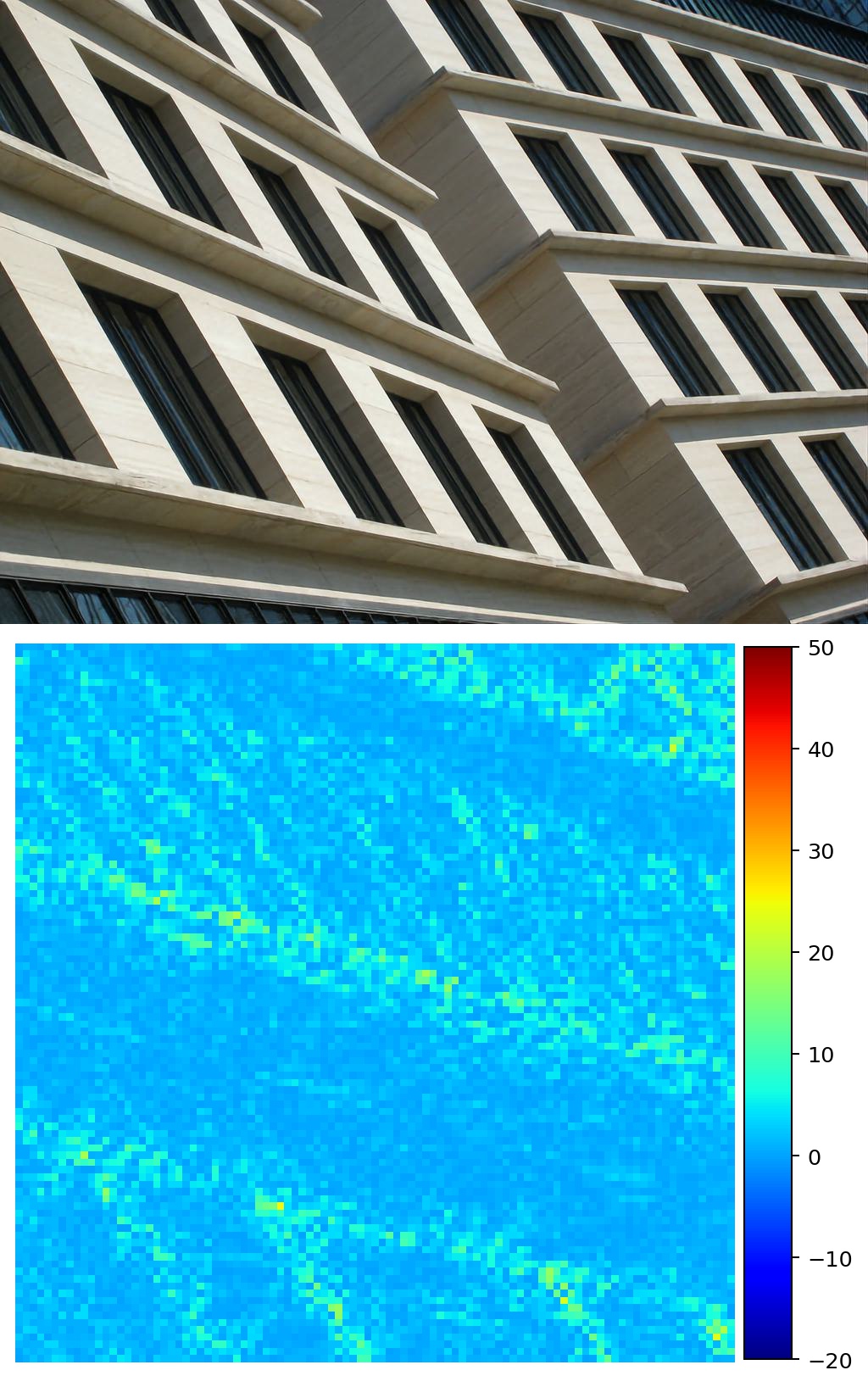}
}

\subfloat[PSNR(dB)/SSIM]{\includegraphics[width=0.78in]{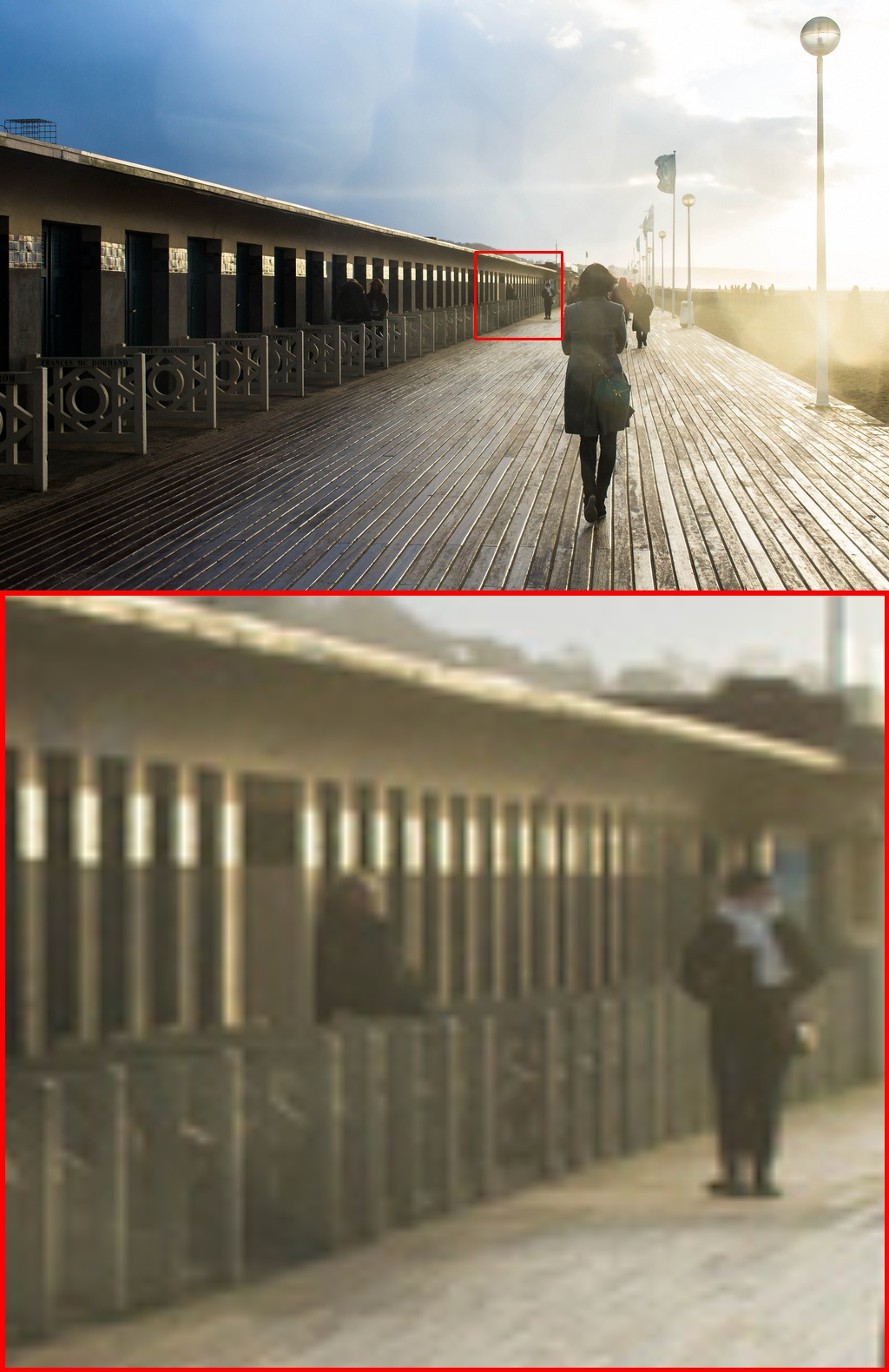}
}
\hfil
\subfloat[28.73/0.8651]{\includegraphics[width=0.78in]{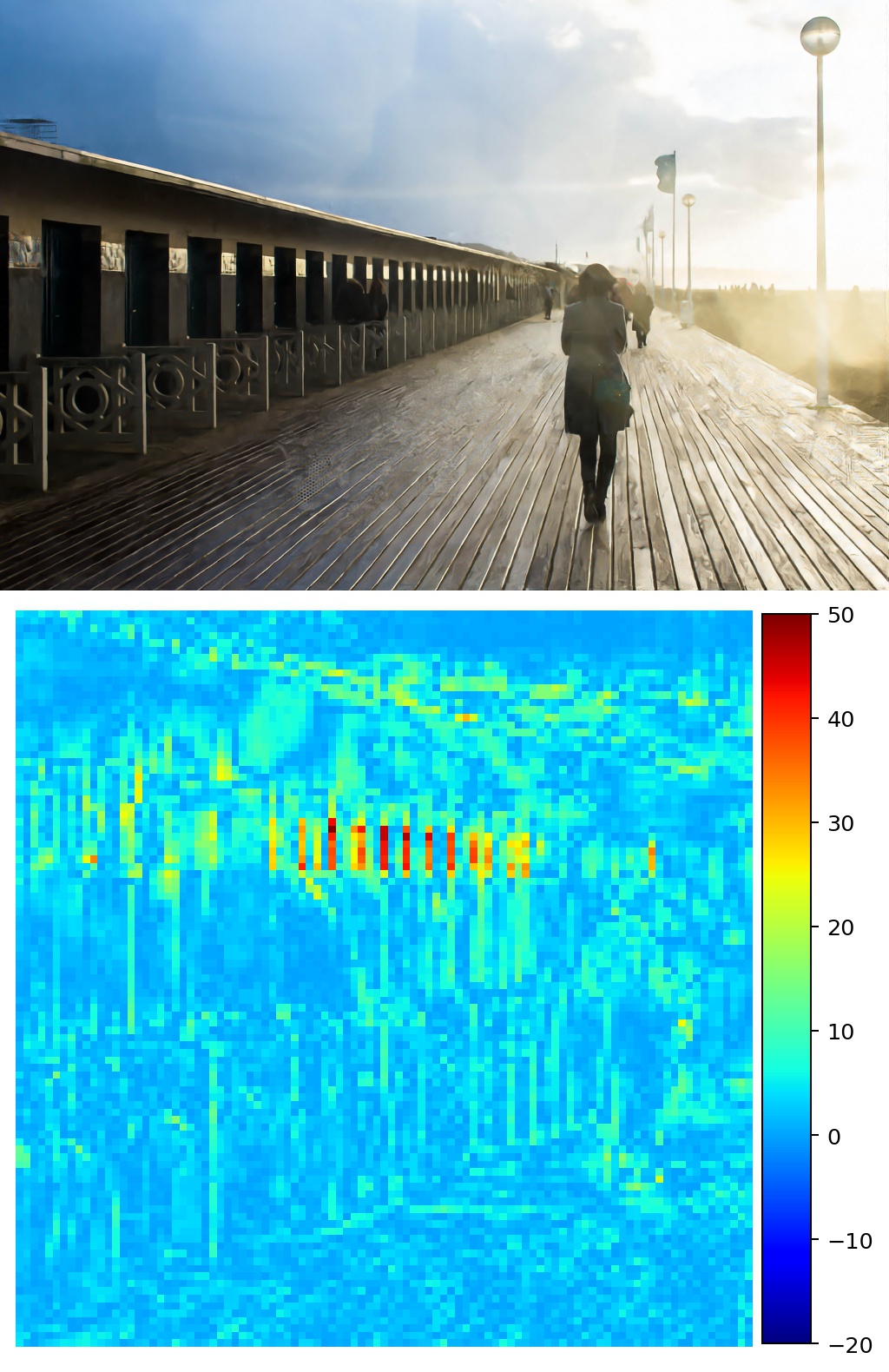}
}
\hfil
\subfloat[27.43/0.8204]{\includegraphics[width=0.78in]{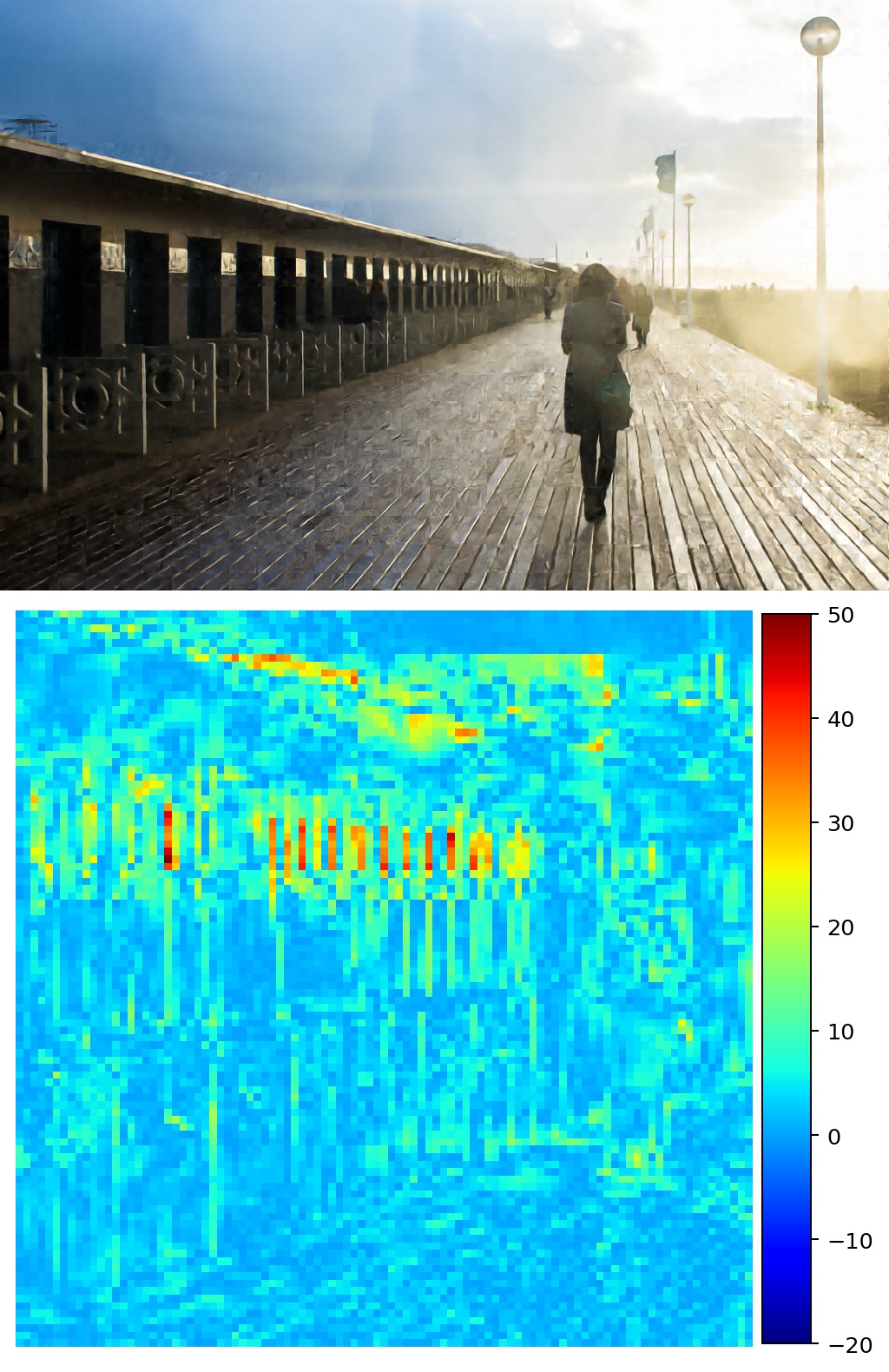}
}
\hfil
\subfloat[\textit{29.19}/\textit{0.8825}]{\includegraphics[width=0.78in]{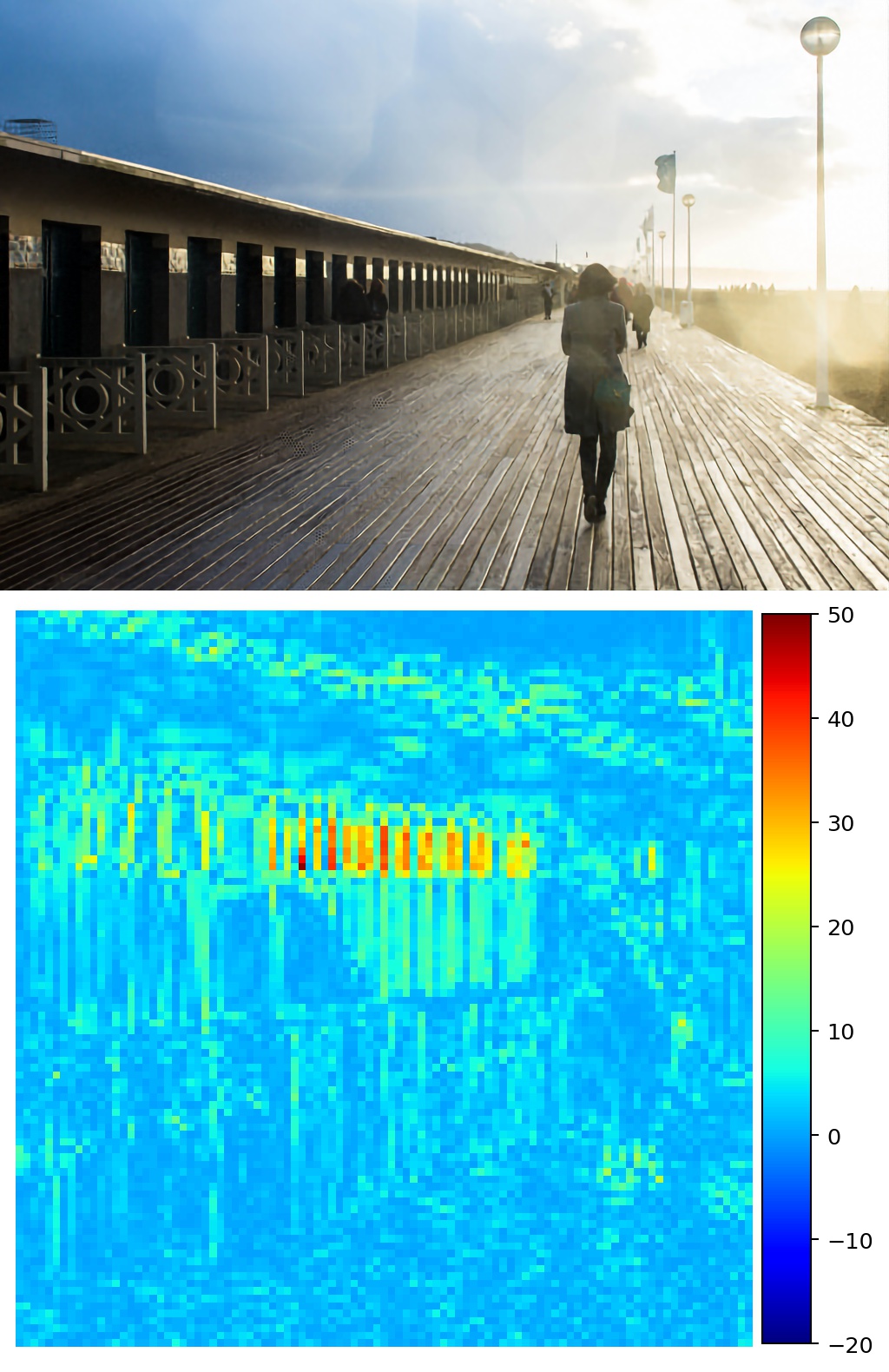}
}
\hfil
\subfloat[28.79/0.8725]{\includegraphics[width=0.78in]{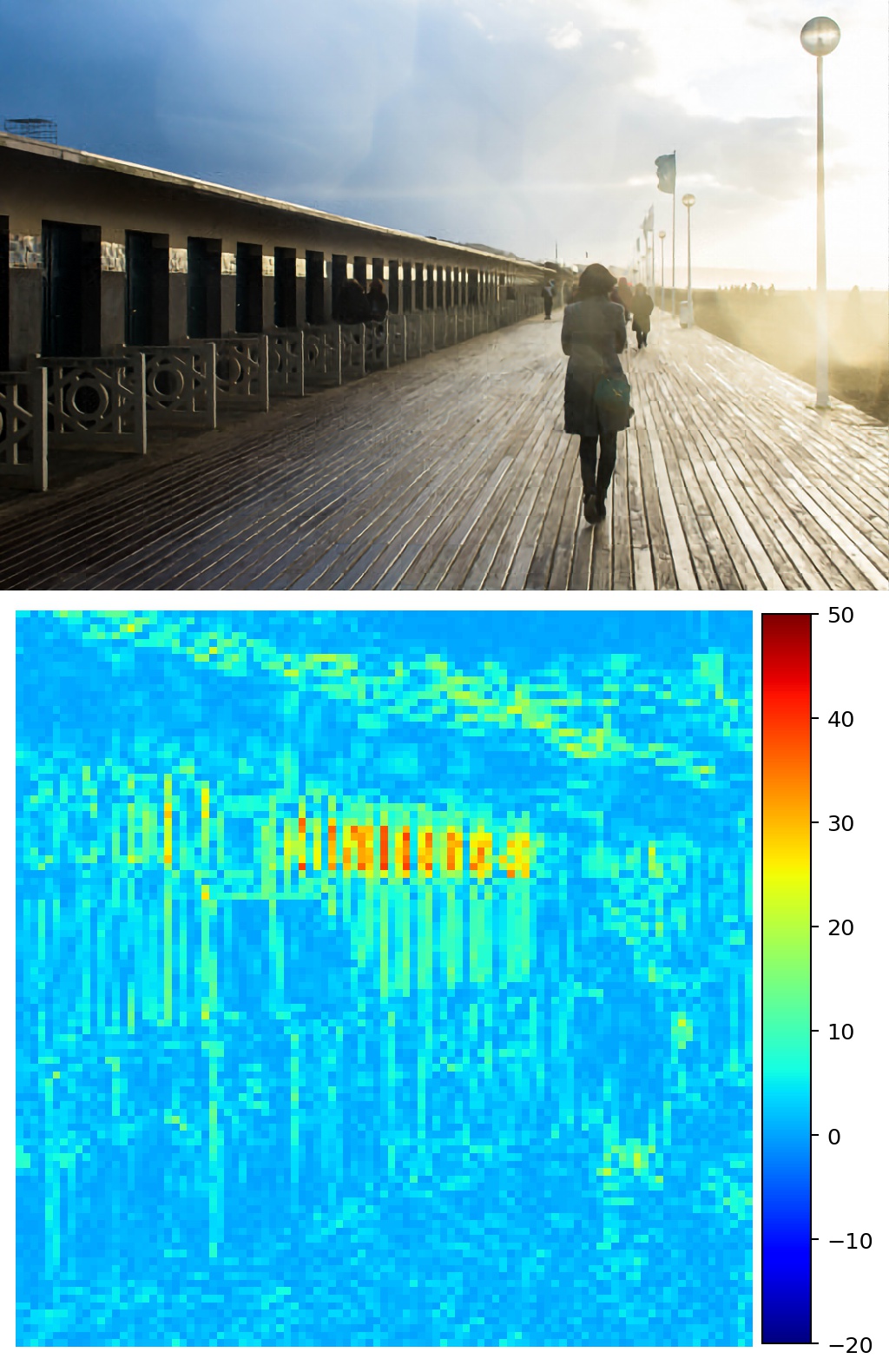}
}
\hfil
\subfloat[28.43/0.8587]{\includegraphics[width=0.78in]{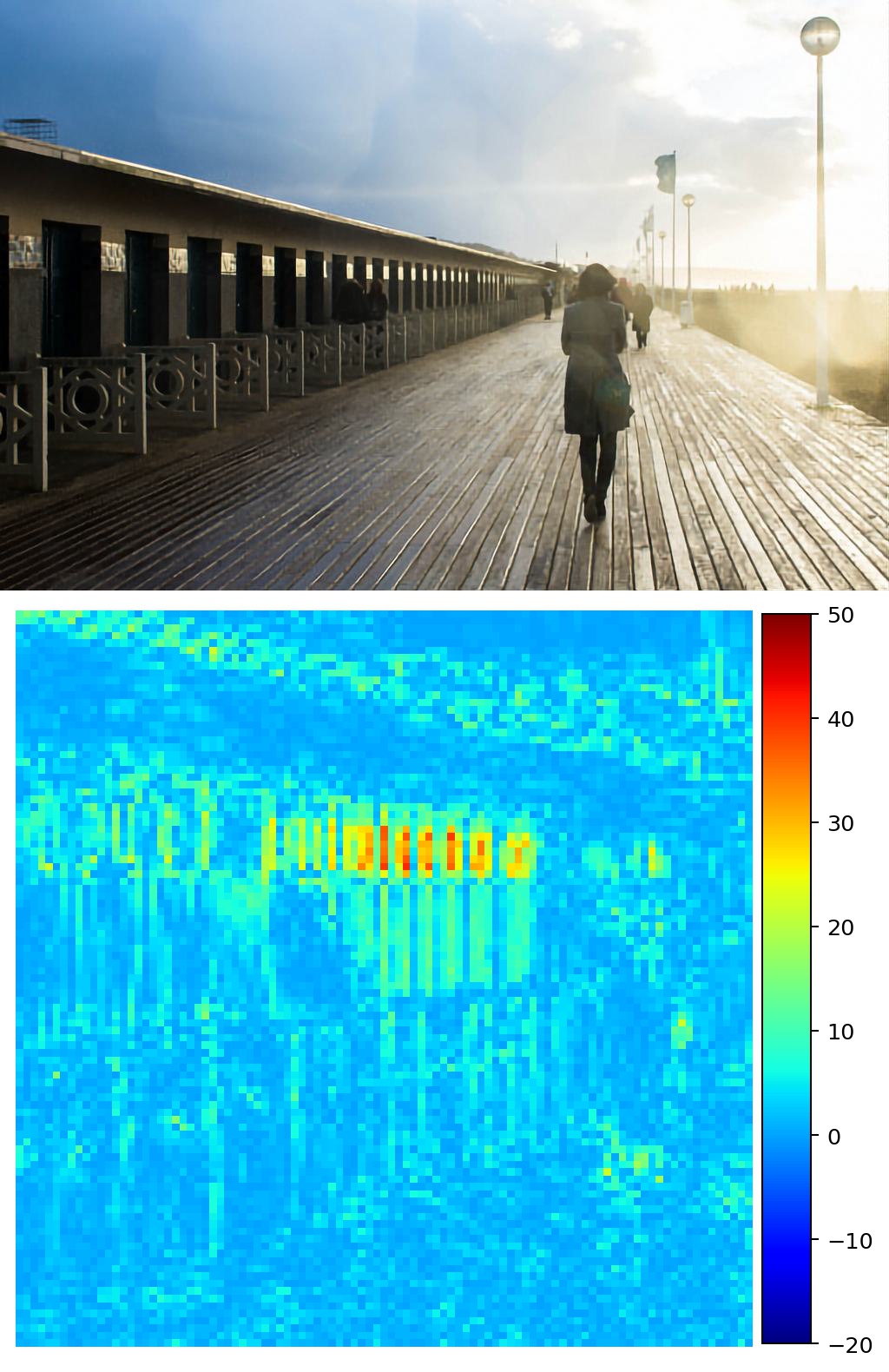}
}
\hfil
\subfloat[\textbf{29.72}/\textbf{0.8874}]{\includegraphics[width=0.78in]{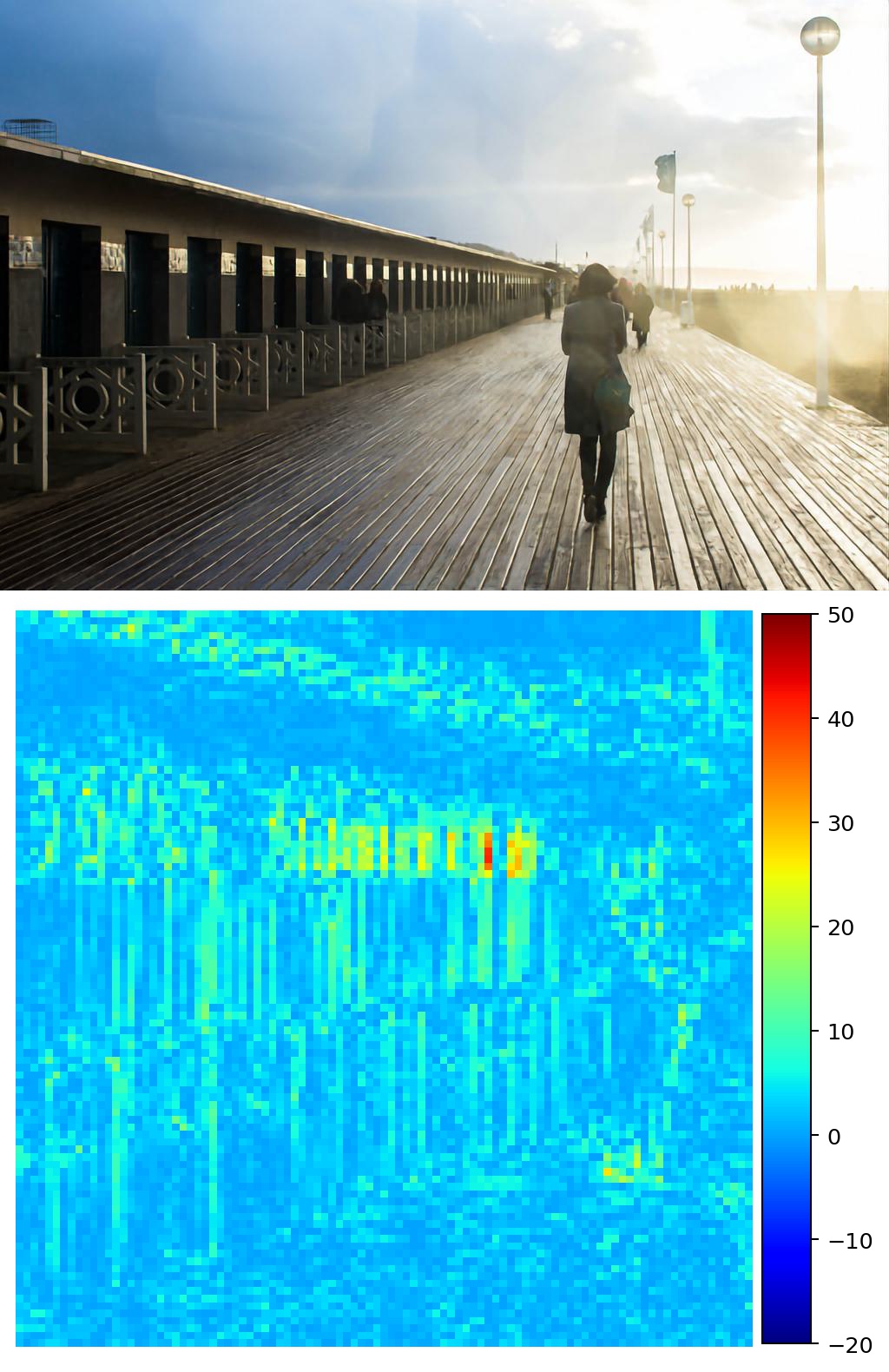}
}
\caption{Visual quality comparisons between the proposed WTDUN and recently state-of-the-art CS methods on Urban100 at 10\% CS ratio. To better illustrate the difference, we show the residual heatmap between the reconstructed image patch and the GT patch. The best and second-best results are highlighted in bold and italics, respectively.}
\label{urban100_img}
\end{figure*}

\textbf{\emph{$\boldsymbol z$-subproblem:}} The goal of the $\boldsymbol z$-subproblem is to generate an intermediate variable to simplify the optimization problem. The $\boldsymbol z$-subproblem is equivalent to solving the following optimization problem:
\begin{equation}
\label{min_z}
\boldsymbol z^k = \mathop{\mathbf{argmin}}\limits_{\boldsymbol z}{\;} \beta \Vert \boldsymbol z\Vert_2 + \frac{\lambda}{2}\Vert \boldsymbol z - \boldsymbol{G{\theta}}^{k-1}\Vert_2^2.
\end{equation}
The mapping process for solving the corresponding optimization $\boldsymbol z$-subproblem is described as follows, which can be solved by using a soft threshold function (STF). It is mapped into a deep network module in our framework. 
\begin{equation}
\label{relu_z}
\boldsymbol z^k = \mathbf{max}(\Vert \boldsymbol{G{\theta}}^{k-1}\Vert_2 - \frac{\beta}{\lambda} , 0)\frac{\boldsymbol{G{\theta}}^{k-1}}{\Vert \boldsymbol{G{\theta}}^{k-1}\Vert_2}.
\end{equation}
We implement the tree-structured group sparsity prior via Eq.~\eqref{relu_z}. Guided by the wavelet tree-structured prior, our unfolding network effectively exploits the inherent structural information embedded in an image. The incorporation of the wavelet tree structure as a guiding principle in our model enables us to exploit the hierarchical nature of wavelet decomposition, facilitating the capture of both local and global features present in an image. By integrating this prior into our framework, we are better equipped to handle complex images characterized by diverse textures and structures. As shown in Fig.~\ref{urban100_img}, it exhibits superior reconstruction performance on the complex dataset Urban100. Moreover, our unfolding network capitalizes on the inter-dependency between subbands at different scales, allowing for leveraging information from one scale to enhance or refine details at another scale. Specifically, the STF consists of the following steps, including:
\begin{itemize}
    \item[(1)]  We concatenate the subbands of different groups in the channel dimension and treat a channel as a group. Then a channel-wise global embedding (CGE) is performed to $wg = \boldsymbol{\theta}^{k-1}$, such as CGE($\boldsymbol{\theta}^{k-1}$). It is composed of channel-wise attention, a ReLU layer, and a $1\times 1$ convolutional.
    \item[(2)] Based on the assumption that a group with higher energy requires a larger threshold, we calculate a group-wise threshold in the shrinkage as follows:
    \begin{equation}
    \boldsymbol{\theta}^{k-1} = wg * \boldsymbol{\theta}^{k-1}.
        \label{relu_z_sub}
    \end{equation}
    \item[(3)]  Applying the group-wise soft-shrinkage operation Eq.~\eqref{relu_z} with the new $\boldsymbol{\theta}^{k-1}$. The soft-shrinkage operation is efficiently implemented by using a ReLU function.
\end{itemize}

\textbf{\emph{$\boldsymbol{\theta}$-subproblem:}} The $\boldsymbol \theta$-subproblem in Eq.~\eqref{min_tree} can be written as follows:
\begin{equation}
\label{min_theta}
\boldsymbol \theta^k = \mathop{\mathbf{argmin}}\limits_{\boldsymbol \theta}{\;} \frac{1}{2}\Vert{\boldsymbol y - \boldsymbol{A\theta}}\Vert_2^2 + \frac{\lambda}{2}\Vert \boldsymbol z^{k} -\boldsymbol{ G\theta}\Vert_2^2 + \beta \Vert \boldsymbol \theta\Vert_1.
\end{equation}
The $\boldsymbol \theta$-subproblem in Eq.~\eqref{min_theta} can be solved through the optimization algorithm of denoising perspective:
\begin{align}
\label{gdm_r}
\boldsymbol{r^{k}} = \mathbf{GDM}(\boldsymbol \theta^{k-1},\boldsymbol z^{k},\lambda)
      =\boldsymbol \theta^{k-1} - \boldsymbol{A}^T(\boldsymbol{A\theta}^{k-1} - \boldsymbol y) - \lambda \boldsymbol{G}^T(\boldsymbol{G\theta}^{k-1} - \boldsymbol z^{k}),
\end{align}
\begin{align}
\label{min_theta1}
\boldsymbol \theta^k = \mathop{\mathbf{argmin}}\limits_{\boldsymbol \theta}{\;} \frac{1}{2}\Vert{\boldsymbol \theta - \boldsymbol r^{k}}\Vert_2^2 + \beta \Vert \boldsymbol \theta\Vert_1,
\end{align}
where GDM denotes the gradient descent module (GDM) at $\boldsymbol r^k$, as shown in Fig.~\ref{frame-stage}. And $\frac{1}{2}\Vert{\boldsymbol y - \boldsymbol{A\theta}}\Vert_2^2 + \frac{\lambda}{2}\Vert \boldsymbol{z}^{k} - \boldsymbol{G\theta}\Vert_2^2 $ is a convex smooth function with Lipschitz constant. The optimization problem in Eq.~\eqref{min_theta1} can be solved as follows:
\begin{align}
\label{relu_theta}
\boldsymbol{\hat \theta}^k &= \mathbf{max}(\Vert \boldsymbol{r}^{k}\Vert_2 - w_i * \beta , 0)\frac{\boldsymbol{r}^{k}}{\Vert \boldsymbol{r}^{k}\Vert_2},
\\
\label{denoise_theta}
\boldsymbol{\theta}^k &= \boldsymbol{\hat \theta}^{k} - (\boldsymbol{A}^T\boldsymbol{A} - \boldsymbol I){\mathfrak{N}}_k(\boldsymbol{\theta}^{k-1}).
\end{align}
We use the denoising block ${\mathfrak{N}}_k(\cdot)$ to remove the noise of $\boldsymbol \theta^{k-1}$ to get the $k$-th reconstruction result $\boldsymbol \theta^k$. Each denoising module consists of four $3 \times 3$ convolutional layers. There is a ReLU activation function between adjacent convolutional layers. A channel attention called GSC in our network is used to learn the corresponding weights $\{w_i|i=1,2,...n\}$ of different image patch groups for improving flexibility.

In each deblocking block $\mathfrak{B}_k(\cdot)$, it consists of six convolutional layers. The ReLU activation function is applied between adjacent convolution layers. By combining $\boldsymbol \theta_{L}$ and $\boldsymbol \theta_{H}$ using the inverse wavelet transform(IDWT), we can obtain the reconstruction result $\boldsymbol X^k$ of the $k$-th phase, which combines the output set $[\boldsymbol X^0, \boldsymbol X^1, ..., \boldsymbol X^{k-1}]$ of the previous $k-1$ phases as input to deblock module. The deblock module integrated inter-context information can effectively remove block artifacts of full-image. We use a cross-domain attention module (CAC) before the inverse wavelet for better collaborative reconstruction of high-frequency and low-frequency components. The process of image deblocking in the $k$-th reconstruction stage can be expressed as:
\begin{subequations}
\begin{align}
\label{deblock_theta}
\boldsymbol X^{k} &= \mathbf{IDWT}(\mathbf{CAC}(\boldsymbol \theta_{L}^k,\boldsymbol \theta_{H}^k)),\\
\boldsymbol X^k_\text{output} &= \boldsymbol X^{k} - {\mathfrak{B}}_k(cat(\boldsymbol X^0, \boldsymbol X^1, ..., \boldsymbol X^{k})).
\end{align}
\end{subequations}
The input of $\mathfrak{B}_k(\cdot)$ is a set of whole concatenated images rather than each image block. Fig.~\ref{block} illustrates the structures of the denoise module, deblock module, and CAC~\cite{ref21} module.
\subsection{Loss Functions}
In this paper, we adopt the mean square error (MSE) as the difference metric between the $m$-th ground-truth $\boldsymbol{X}_\text{m} $ in the training set and the final recovered result $\boldsymbol X_\text{m}^K$. The loss function is formulated as:
\begin{equation}
\label{loss_mse}
{\mathcal{L}}_\text{MSE} = \frac{1}{N_\text{a} N_\text{b}}\Vert{\widetilde{\boldsymbol X}_\text{m} - \boldsymbol X_\text{m}^K}\Vert_2^2,
\end{equation}
where $N_\text{a}$ denotes the size of  $\widetilde{\boldsymbol{X}}_\text{m}$ and $N_\text{b}$ denotes the size of the training set.

The high-frequency coefficients potentially converge to zero by performing convolution operations on them. To address this issue, we develop a texture loss function that aims to prevent the high-frequency coefficients from converging to zero:
\begin{equation}
\label{loss_dwttexture}
{\mathcal{L}}_\text{texture} = \mathbf{max}(\Vert{\boldsymbol{\widetilde \theta}_\text{H}}\Vert - \Vert{\boldsymbol{\theta}_\text{H}^k}\Vert + \epsilon , 0),
\end{equation}
where $\boldsymbol{\widetilde \theta}_\text{H}$ is the high-frequency coefficients of the original image in the training set, $ \boldsymbol{\theta}_\text{H}^k$ is the reconstruction result at $k$-th stage. $\epsilon$ is a bias term that prevents the texture from disappearing, which is a tensor of the same size as the high-frequency subbands and has values only at non-zero elements.

The initial loss denoted as the mean square error (MSE) between the output result $\boldsymbol{X}_\text{init}$ of the initial reconstruction module and the corresponding  original image in the training set, defined as:
\begin{equation}
\label{loss_dwt}
{\mathcal{L}}_\text{init} = \frac{1}{N_\text{a} N_\text{b}}\Vert{\boldsymbol{\widetilde{X}}_\text{m} - \boldsymbol{X}_\text{init}}\Vert_2^2.
\end{equation}
Finally, an end-to-end loss function is designed as follows:
\begin{equation}
\label{loss_total}
{\mathcal{L}}_\text{total} = {\mathcal{L}}_\text{MSE} + \gamma * {\mathcal{L}}_\text{texture} + \mu * {\mathcal{L}}_\text{init},
\end{equation}
where $\gamma$, $\mu$ are the regularization parameters. In our experiments, $\gamma$ and $\mu$ are set to $0.001$ and $0.01$ respectively.

\begin{table*}[!t]
\small
\centering
\caption{Average PSNR(dB)/SSIM performance comparisons on Set5 with  different CS sampling rates. * indicates the case where the result at the corresponding sampling rate is missing in the original paper. The best and second-best results are highlighted in bold and italics, respectively.}
\begin{tabular}{c|c|c|c|c|c} 
\toprule
\multirow{2}{*}{METHOD} & \multicolumn{5}{c}{Set5 \qquad PSNR(dB)/SSIM}   \\
\cline{2-6}  
  & {10\%} & {20\%} & {30\%} & {40\%} & {50\%}  \\
\toprule
  DPA-Net\cite{ref1} & 30.32/0.8713  & */* & 36.17/0.9495  & 38.05/0.9632  & 39.57/0.9716  \\ 
  AMP-Net\cite{ref3}   & {31.95}/0.9017  &\textit{35.49}/{0.9419} & \textit{37.86}/0.9606  & {39.70}/0.9713  & 41.51/0.9791  \\   
  COAST\cite{ref5} & 30.50/0.8794 & 34.18/0.9298 & 36.48/0.9515 & 38.33/0.9645  & 40.21/0.9744  \\ 
  ISTA-Net++\cite{ref6}  & 29.61/0.8563 & 33.33/0.9173 & 35.62/0.9427  &  37.40/0.9575  & 38.94/0.9678  \\ 
  MADUN\cite{ref7}  & 31.11/0.8910  & 34.80/0.9363 & 37.25/0.9561  & 39.29/0.9693  & 41.18/0.9784  \\ 
  ULAMP\cite{ref10} & 30.78/0.8774  & 31.94/0.8868 & 36.39/0.9599  & 38.20/0.9693  & 40.97/\textit{0.9827}  \\ 
  DPUNet\cite{ref11} & 31.80/{0.9079}  & 35.38/\textit{0.9458} & 37.54/0.9618  & 39.44/0.9716  &  41.10/0.9783  \\ 
  Ours(64$\times$64) &  \textit{32.14}/\textit{0.9123} & 35.39/0.9443 &  37.82/\textit{0.9622} &  \textit{39.94}/\textit{0.9734} & \textit{42.09}/0.9824 \\
  Ours(128$\times$128) &  \textbf{32.23}/\textbf{0.9132} &  \textbf{35.56}/\textbf{0.9461} &  \textbf{38.04}/\textbf{0.9637}  &  \textbf{40.14}/\textbf{0.9746}  &  \textbf{42.18}/\textbf{0.9827}  \\
  \bottomrule
  \end{tabular}
\label{tb1}
\end{table*}
\begin{table*}[!t]
\small
\centering
\caption{Average PSNR(dB)/SSIM performance comparisons on Set11 with  different CS sampling rates. * indicates the case where the result at the corresponding sampling rate is missing in the original paper. The best and second-best results are highlighted in bold and italics, respectively.}
\begin{tabular}{c|c|c|c|c|c} 
\toprule
\multirow{2}{*}{METHOD} & \multicolumn{5}{c}{Set11 PSNR(dB)/SSIM}   \\
\cline{2-6}  
    & {10\%} & {20\%} & {30\%} & {40\%} & {50\%}  \\
\toprule
  DPA-Net\cite{ref1}  & 26.99/0.8354  & */* & */*  & 35.04/0.9565  & 36.73/0.9670  \\ 
  BCS-Net\cite{ref4}  & 29.36/0.8650  & 32.87/0.9254 & 35.40/0.9527  & 36.52/0.9640  & 39.58/0.9734  \\ 
  COAST\cite{ref5}  & 28.69/0.8618  & 32.54/0.9251 & 35.04/0.9501  & 37.13/0.9648  & 38.94/0.9744  \\ 
  ISTA-Net++\cite{ref6}   & 27.62/0.8358  & 31.66/0.9127 &  34.23/0.9427  & 36.28/0.9593  & 37.94/0.9693  \\ 
  MADUN\cite{ref7}  & 29.29/0.8768  & 33.30/0.9355 & 36.00/0.9576  & {38.09}/0.9700  & {39.86}/0.9774  \\ 
  FHDUN\cite{ref9} & 29.53/0.8859  & */* & 36.12/\textit{0.9589}  & 38.04/0.9696  & */*  \\ 
  DPUNet\cite{ref11} & 29.30/0.8815  & 33.17/0.9357 & 35.75/0.9581 & 37.90/\textit{0.9705}  & 39.69/0.9782  \\ 
   SODAS-Net\cite{ref42} & 28.89/0.8669  & 32.20/0.9243  & 35.55/0.9543  & 37.74/0.9680  &  39.60/0.9769  \\ 
   DPC-DUN\cite{ref46} & 29.40/0.8798  & 33.10/0.9334 & 35.88/0.9570  & 37.98/0.9694  &  39.84/0.9778  \\ 
  Ours(64$\times$64)     &\textit{29.53}/\textit{0.8867} &\textit{33.58}/\textit{0.9371} & \textit{36.27}/0.9585 & \textit{38.26}/0.9698 & \textit{40.22}/\textit{0.9784}  \\
  Ours(128$\times$128) &  \textbf{29.64}/\textbf{0.8877}  &  \textbf{33.81}/\textbf{0.9399} &  \textbf{36.41}/\textbf{0.9597}  &  \textbf{38.45}/\textbf{0.9708}  &  \textbf{40.38}/\textbf{0.9789}
  \\
  \bottomrule
  \end{tabular}
\label{tb2}
\end{table*}
\begin{table*}[!t]
\small
\centering
\caption{Average PSNR(dB)/SSIM performance comparisons on Set14 with different CS sampling rates. The best and second-best results are highlighted in bold and italics, respectively.}
\begin{tabular}{c|c|c|c|c|c} 
\toprule
\multirow{2}{*}{METHOD} & \multicolumn{5}{c}{Set14 PSNR(dB)/SSIM}   \\
\cline{2-6}  
   & {10\%} & {20\%} & {30\%} & {40\%} & {50\%}  \\
\toprule
  AMP-Net\cite{ref3} & \textit{28.69}/0.8171  & \textit{31.95}/0.8933 & 34.27/0.9293  & 36.26/\textit{0.9505}  &  38.10/\textit{0.9647}  \\
  COAST\cite{ref5} & 27.41/0.7799  & 30.71/0.8672 & 33.10/0.9106  & 35.12/0.9369  & 36.93/0.9549  \\ 
  ISTA-Net++\cite{ref6} & 26.75/0.7549  & 30.09/0.8518 &  32.40/0.8999  & 34.26/0.9287  & 35.90/0.9477  \\ 
  MADUN\cite{ref7} & 27.97/0.7914  & 31.50/0.8790 & 34.05/0.9194  & 36.05/0.9439  & 37.96/0.9592  \\ 
  DPUNet\cite{ref11}  & 28.49/\textit{0.8226} & 31.61/\textbf{0.8963} & 33.95/\textit{0.9308} & 35.93/0.9505  & 37.66/0.9628  \\ 
  Ours(64$\times$64)     &28.61/0.8216  &31.89/0.8908 & \textit{34.36}/0.9280  & \textit{36.27}/0.9486  & \textit{38.31}/0.9644 \\
  Ours(128$\times$128) & \textbf{28.71}/\textbf{0.8235}  & \textbf{32.10}/\textit{0.8944} & \textbf{34.57}/\textbf{0.9308}  & \textbf{36.53}/\textbf{0.9507} & \textbf{38.49}/\textbf{0.9653}
  \\
  \bottomrule
  \end{tabular}
\label{tb3}
\end{table*}
\begin{table*}[!t]
\small
\centering
\caption{Average PSNR(dB)/SSIM performance comparisons on Urban100 with  different CS sampling rates. * indicates the case where the result at the corresponding sampling rate is missing in the original paper. The best and second-best results are highlighted in bold and italics, respectively.}
\begin{tabular}{c|c|c|c|c|c} 
\toprule
\multirow{2}{*}{METHOD} & \multicolumn{5}{c}{Urban100 PSNR(dB)/SSIM}   \\
\cline{2-6}  
    & {10\%} & {20\%} & {30\%} & {40\%} & {50\%}  \\
\toprule
  DPA-Net\cite{ref1}  & 24.55/0.7841  & */* &  29.47/0.9034  & 31.09/0.9311  & 32.08/0.9447  \\ 
  OPINE-Net\cite{ref2} & 26.56/0.8345  & 30.07/0.9088 &32.64/0.9419  & 34.66/0.9600  & 36.64/0.9727  \\ 
  AMP-Net\cite{ref3}& 25.96/0.8133  & 29.50/0.8974 & 32.07/0.9352  & 34.22/0.9569  & 36.16/0.9706  \\ 
  COAST\cite{ref5} & 25.94/0.8038  & 29.70/0.8940 & 32.20/0.9317  & 34.21/0.9528  & 35.99/0.9665  \\ 
  ISTA-Net++\cite{ref6} &  24.78/0.7607  & 28.55/0.8687 & 31.08/0.9152  & 33.10/0.9402  & 34.86/0.9560  \\ 
  DPUNet\cite{ref11}   & 26.10/0.8226  & 29.71/0.9027 &  32.23/0.9378 &34.30/0.9573  & 36.10/0.9693  \\ 
   SODAS-Net\cite{ref42}  & 26.23/0.8084  & 29.51/0.8950 & 33.15/0.9412  & 35.28/0.9599  &  37.14/0.9721  \\ 
  Ours(64$\times$64)    &  \textit{26.58}/\textit{0.8358} & \textit{30.58}/\textit{0.9109} & \textit{33.18}/\textit{0.9431} & 
  \textit{35.26}/\textit{0.9610} & \textit{37.39}/\textit{0.9738} \\
  Ours(128$\times$128)  & \textbf{26.77}/\textbf{0.8402}  & \textbf{30.82}/\textbf{0.9142 }& \textbf{33.44}/\textbf{0.9459}  & \textbf{35.60}/\textbf{0.9629}  & \textbf{37.65}/\textbf{0.9748}  \\
  \bottomrule
  \end{tabular}
\label{tb4}
\end{table*}
\begin{figure*}[t]
\captionsetup[subfigure]{labelformat=empty}
\centering
GT \ \quad COAST\cite{ref5}  \ ISTA-Net++\cite{ref6}  \ MADUN\cite{ref7}   \ DPUNet\cite{ref11} \ AMP-Net\cite{ref3} \quad Ours  \qquad \
\\
\subfloat[PSNR(dB)/SSIM]{\includegraphics[width=0.77in]{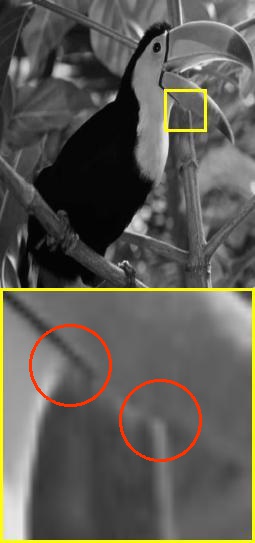}
}
\hfil
\subfloat[34.07/0.9399]{\includegraphics[width=0.77in]{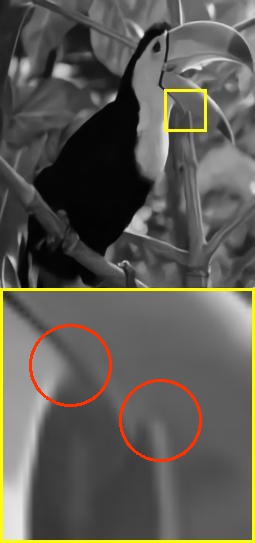}
}
\hfil
\subfloat[32.31/0.9098]{\includegraphics[width=0.77in]{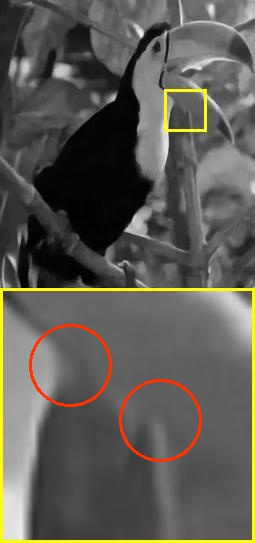}
}
\hfil
\subfloat[35.06/0.9509]{\includegraphics[width=0.77in]{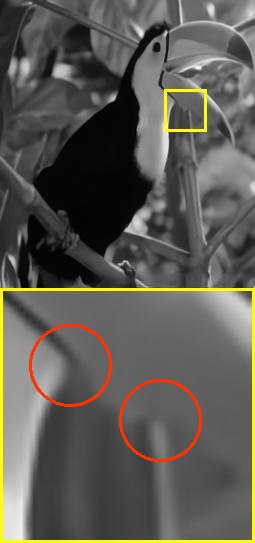}
}
\hfil
\subfloat[34.92/\textit{0.9566}]{\includegraphics[width=0.77in]{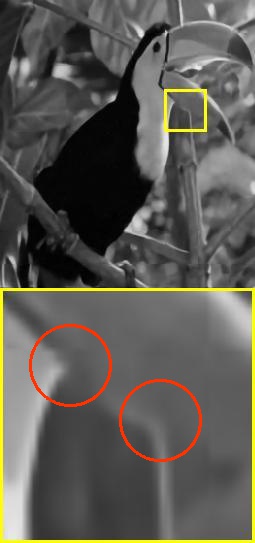}
}
\hfil
\subfloat[{\textit{35.11}}/0.9535]{\includegraphics[width=0.77in]{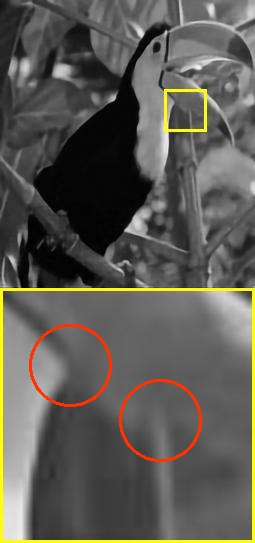}
}
\hfil
\subfloat[{\textbf{35.59}/\textbf{0.9624}}]{\includegraphics[width=0.77in]{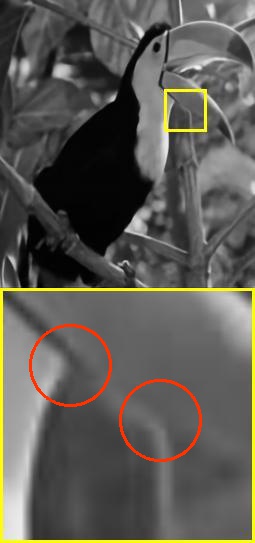}
}

\subfloat[PSNR(dB)/SSIM]{\includegraphics[width=0.77in]{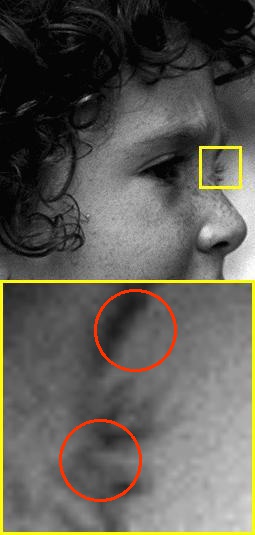}
}
\hfil
\subfloat[30.93/0.7660]{\includegraphics[width=0.77in]{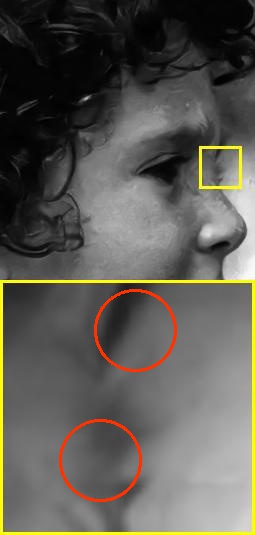}
}
\hfil
\subfloat[30.44/0.7443]{\includegraphics[width=0.77in]{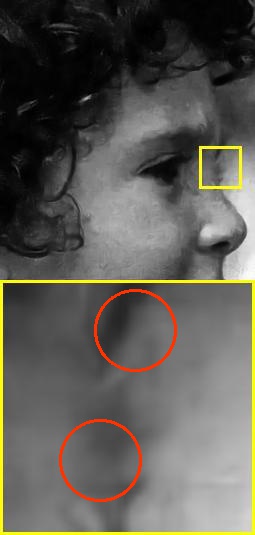}
}
\hfil
\subfloat[30.99/0.7731]{\includegraphics[width=0.77in]{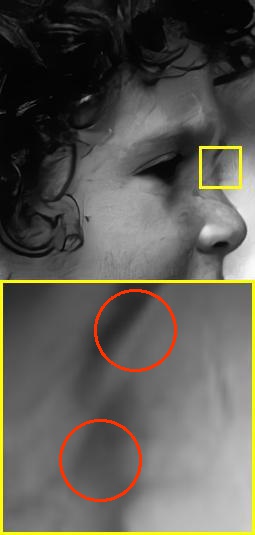}
}
\hfil
\subfloat[\textit{32.17}/\textit{0.8089}]{\includegraphics[width=0.77in]{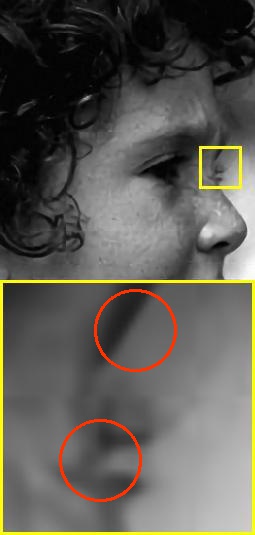}
}
\hfil
\subfloat[32.09/0.7898]{\includegraphics[width=0.77in]{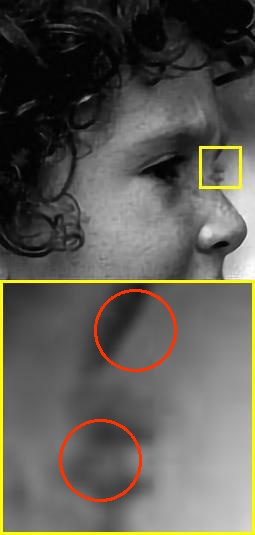}
}
\hfil
\subfloat[{\textbf{32.31}/\textbf{0.8143}}]{\includegraphics[width=0.77in]{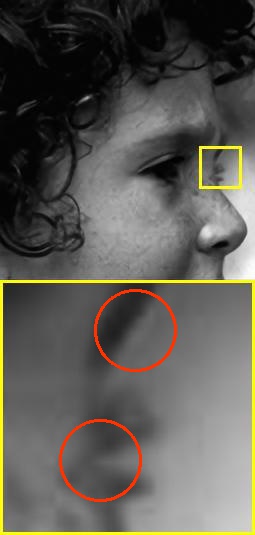}
}
\caption{Visual quality comparisons between our proposed method and recently state-of-the-art CS methods on Set5 at 10\% CS ratio. The best and second-best results are highlighted in bold and italics, respectively.}
\label{set5_img}
\vspace{-0.2cm}
\end{figure*}
\begin{figure*}[t]
\centering
\captionsetup[subfigure]{labelformat=empty}
GT \ \quad COAST\cite{ref5}  \ ISTA-Net++\cite{ref6}  \ MADUN\cite{ref7}   \ DPUNet\cite{ref11} \ AMP-Net\cite{ref3} \quad Ours  \qquad \
\\
\subfloat[PSNR(dB)/SSIM]{\includegraphics[width=0.77in]{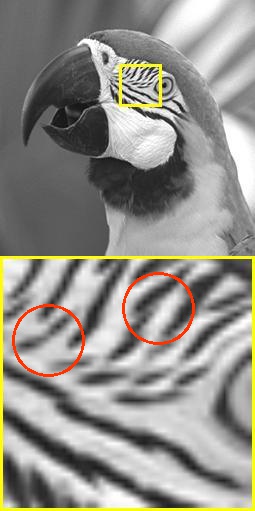}
}
\hfil
\subfloat[28.30/0.8929]{\includegraphics[width=0.77in]{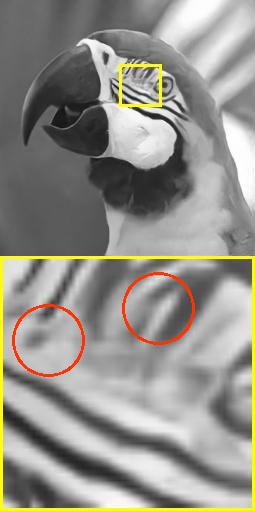}
}
\hfil
\subfloat[27.65/0.8743]{\includegraphics[width=0.77in]{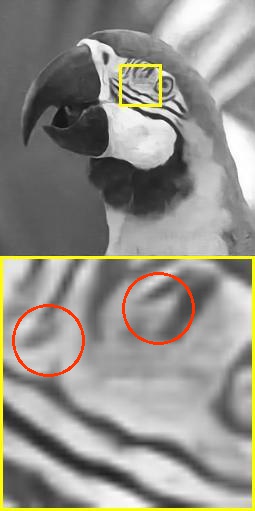}
}
\hfil
\subfloat[28.83/0.9000]{\includegraphics[width=0.77in]{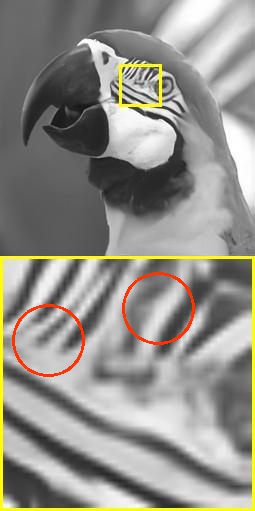}
}
\hfil
\subfloat[28.76/\textit{0.9089}]{\includegraphics[width=0.77in]{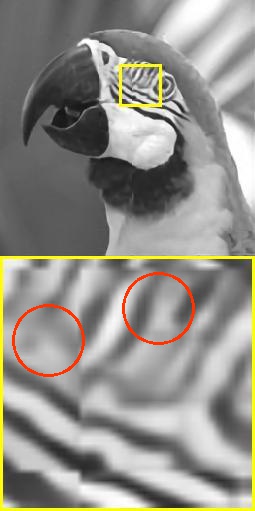}
}
\hfil
\subfloat[\textit{29.02}/0.9038]{\includegraphics[width=0.77in]{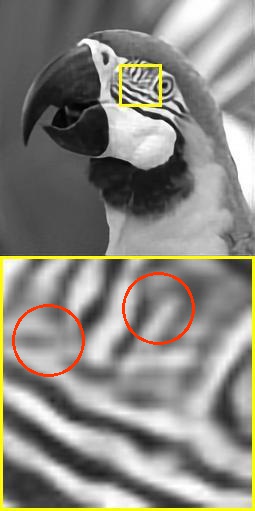}
}
\hfil
\subfloat[{2\textbf{9.58}/\textbf{0.9182}}]{\includegraphics[width=0.77in]{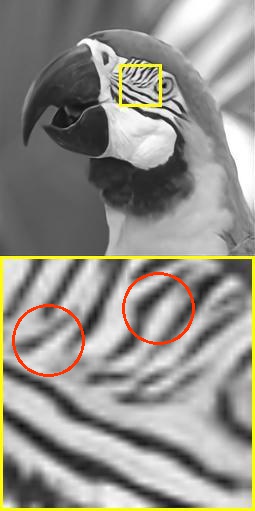}
} 
\hfil
\subfloat[PSNR(dB)/SSIM]{\includegraphics[width=0.77in]{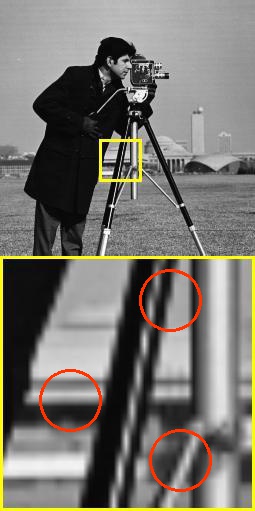}
}
\hfil
\subfloat[26.11/0.8245]{\includegraphics[width=0.77in]{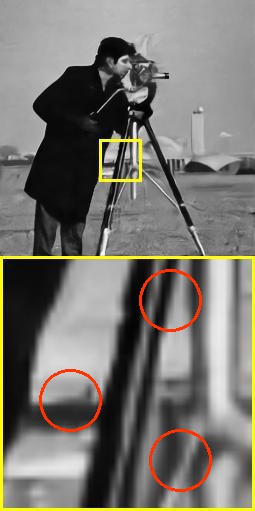}
}
\hfil
\subfloat[25.09/0.8031]{\includegraphics[width=0.77in]{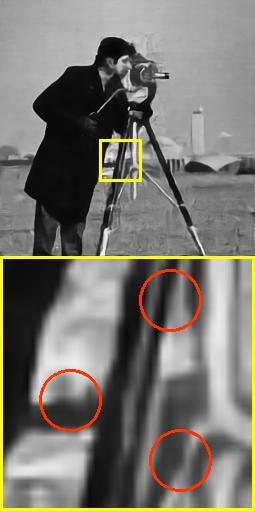}
}
\hfil
\subfloat[26.41/0.8344]{\includegraphics[width=0.77in]{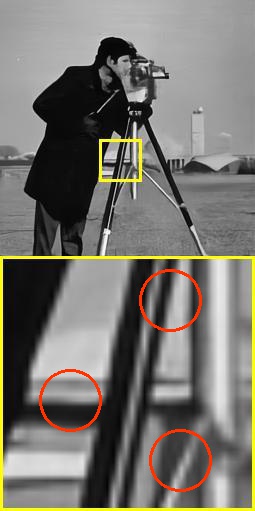}
}
\hfil
\subfloat[26.40/0.8477]{\includegraphics[width=0.77in]{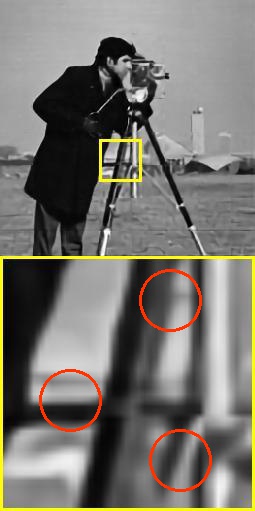}
}
\hfil
\subfloat[\textit{26.59}/\textbf{0.8528}]{\includegraphics[width=0.77in]{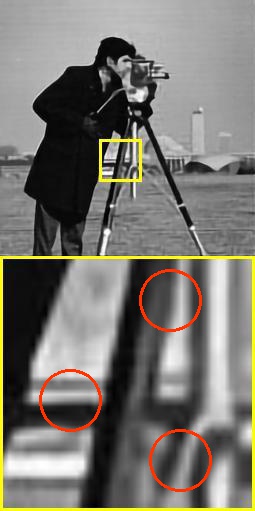}
}
\hfil
\subfloat[\textbf{27.18}/\textit{0.8513}]{\includegraphics[width=0.77in]{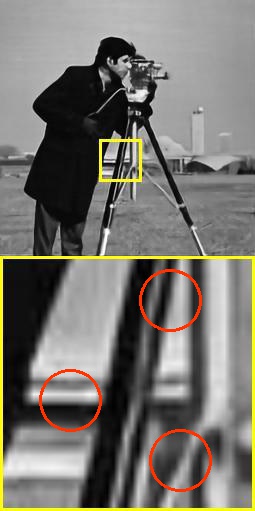}
}
\caption{Visual quality comparisons between our WTDUN and recently state-of-the-art CS methods on Set11 at 10\% CS ratio. The best and second-best results are highlighted in bold and italics, respectively.}
\label{set11_img}
\vspace{-0.2cm}
\end{figure*}
\begin{figure*}[!t]
\centering
\captionsetup[subfigure]{labelformat=empty}
GT \ \quad COAST\cite{ref5}  \ ISTA-Net++\cite{ref6}  \ MADUN\cite{ref7}   \ DPUNet\cite{ref11} \ AMP-Net\cite{ref3} \quad Ours  \qquad \
\\
\subfloat[PSNR(dB)/SSIM]{\includegraphics[width=0.77in]{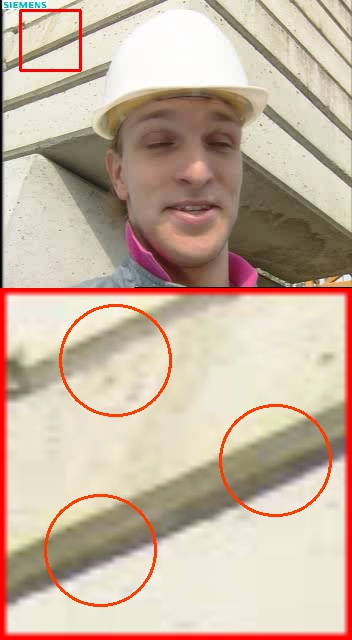}
}
\hfil
\subfloat[32.78/0.9207]{\includegraphics[width=0.77in]{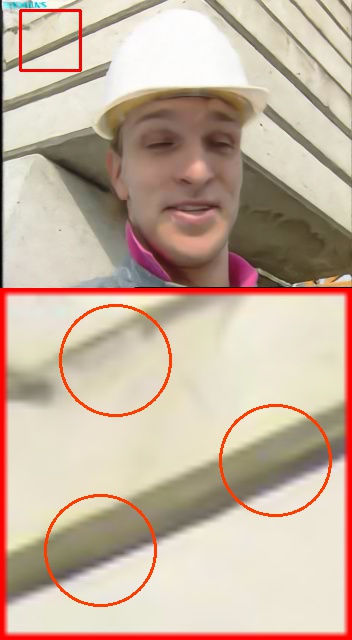}
}
\hfil
\subfloat[31.27/0.9028]{\includegraphics[width=0.77in]{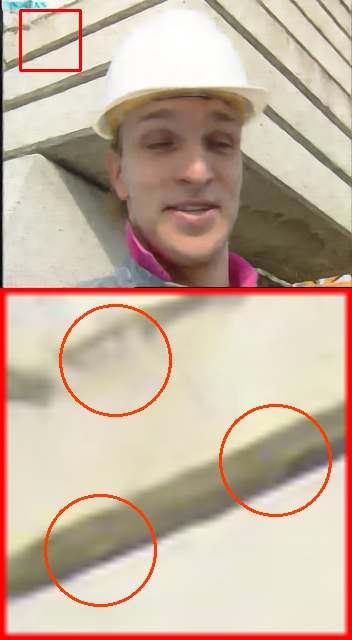}
}
\hfil
\subfloat[\textit{33.70}/0.9256]{\includegraphics[width=0.77in]{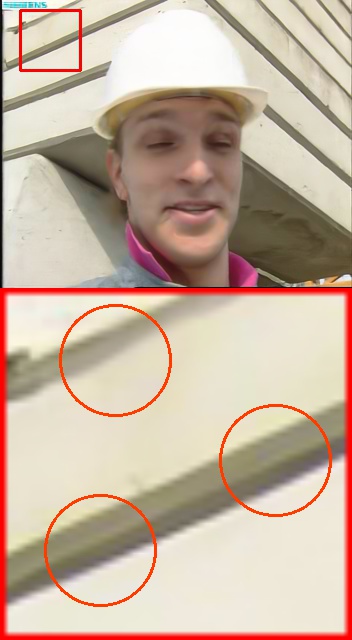}
}
\hfil
\subfloat[33.07/\textit{0.9304}]{\includegraphics[width=0.77in]{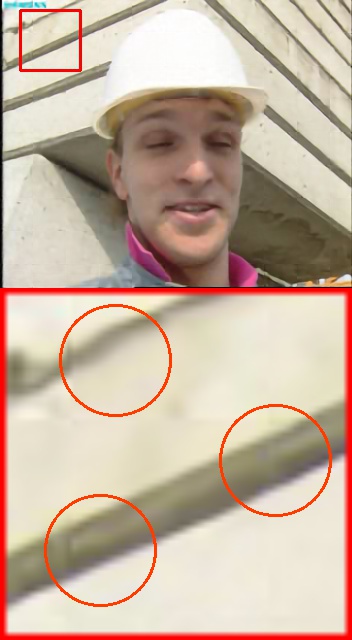}
}
\hfil
\subfloat[32.54/0.9296]{\includegraphics[width=0.77in]{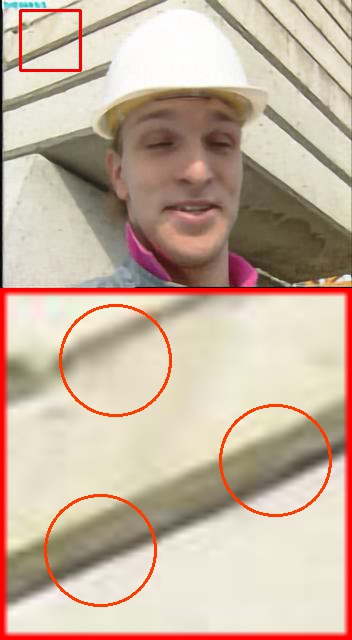}
}
\hfil
\subfloat[\textbf{33.85}/\textbf{0.9381}]{\includegraphics[width=0.77in]{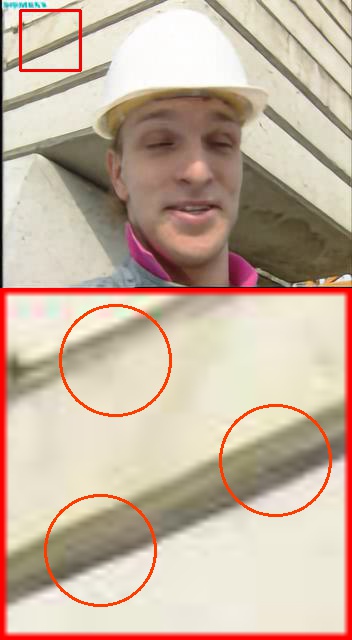}
}
\caption{Visual quality comparisons between our WTDUN and recently state-of-the-art CS methods on Set14 at 10\% CS ratio. The best and second-best results are highlighted in bold and italics, respectively.}
\label{set14_img}
\vspace{-0.55cm}
\end{figure*}
\begin{table*}[!ht]
\centering
\caption{{Average PSNR/SSIM comparisons with other multi-scale sampling methods on Set11 dataset. The best and second-best results are highlighted in bold and italics, respectively.}}
\small
\begin{tabular}{c|c|c|c|c|c} 
\toprule
\multirow{2}{*}{~Method} & \multicolumn{5}{c}{Sampling Rate}  \\
\cline{2-6}  
&  10\% & 20\%  & 30\% &40\% &50\% \\
\toprule
BCS-SPL-DWT\cite{ref016}   & 23.31/0.7056
  & 27.68/0.8057 & 28.41/0.8614  & 30.32/0.8977  & 32.13/0.9241 \\ 
MS-BCS-SPL\cite{ref015}&  27.00/0.8293 & 30.21/0.8893& 32.51/0.9398 & 34.01/0.9428 & 35.33/0.9489
  \\
MS-DCSNet\cite{ref51} & 28.58/0.8648 & 32.20/0.9215 & 34.63/0.9478 &36.65/0.9627&  38.68/0.9740
\\
Ours(64$\times$64)     &\textit{29.53}/\textit{0.8867} &\textit{33.58}/\textit{0.9371} & \textit{36.27}/\textit{0.9585} & \textit{38.26}/\textit{0.9698} & \textit{40.22}/\textit{0.9784}  \\
  Ours(128$\times$128) &  \textbf{29.64}/\textbf{0.8877}  &  \textbf{33.81}/\textbf{0.9399} &  \textbf{36.41}/\textbf{0.9597}  &  \textbf{38.45}/\textbf{0.9708}  &  \textbf{40.38}/\textbf{0.9789}
\\
  \bottomrule
  \end{tabular}
\label{ms_sample}
\vspace{-0.15cm}
\end{table*}
\begin{table*}[!t]
\small
\centering
\caption{The effect of wavelet domain adaptive sampling method (WAS), wavelet tree-structure prior (WTP), deblock, memory context (MC) module, and cross-domain attention (CAC) on reconstruction quality. The sampling ratio of CS measurement is 20\% and 50\% on the Urban100 dataset. The best results are highlighted in bold.}
\begin{tabular}{c|c|c|c|c|c|c|c} 
\toprule
\multirow{2}{*}{Cases} & \multicolumn{5}{c}{Modules} & \multicolumn{2}{c}{PSNR/SSIM}\\
\cline{2-8}  
   & { WAS} & {WTP} & {Deblock} & {MC} & {CAC} & {20\%} & {50\%} \\
\toprule
  (a)          &  & \checkmark  & \checkmark & \checkmark & \checkmark & 28.59/0.8649 &  35.61/0.9600
\\
  (b)          & \checkmark &   & \checkmark & \checkmark  & \checkmark & 30.45/0.9010 &  37.38/0.9710 
\\
  (c)          & \checkmark &  \checkmark &  &  \checkmark& \checkmark  & 28.05/0.8588 &  34.43/0.9552 
  \\
 (d)          & \checkmark & \checkmark  & \checkmark &  & \checkmark & 30.30/0.8995 & 37.12/0.9703
\\
(e)          & \checkmark & \checkmark  & \checkmark & \checkmark &  & 30.31/0.9010 & 36.97/0.9699
\\
  WTDUN          & \checkmark  & \checkmark  & \checkmark & \checkmark & \checkmark & \textbf{30.58}/\textbf{0.9109} & \textbf{37.39}/\textbf{0.9738}
  \\
  \bottomrule
  \end{tabular}
\label{ablation100}
\end{table*}
\begin{table*}[!t]
\small
\centering
\caption{The effect of wavelet domain adaptive sampling method (WAS), wavelet tree-structure prior (WTP), deblock, memory context (MC) module, and cross-domain attention (CAC) on reconstruction quality. The sampling ratio of CS is 20\% and 50\% on the Set11 dataset. The best results are highlighted in bold.}
\begin{tabular}{c|c|c|c|c|c|c|c} 
\toprule
\multirow{2}{*}{Cases} & \multicolumn{5}{c}{Modules} & \multicolumn{2}{c}{PSNR/SSIM}   \\
\cline{2-8}  
   & { WAS} & {WTP} & {Deblock} & {MC}& {CAC} & {20\%}  & {50\%}\\
\toprule
  (a)          &  & \checkmark  & \checkmark & \checkmark & \checkmark  & 31.63/0.9080 & 38.02/0.9675
\\
  (b)          & \checkmark &   & \checkmark & \checkmark & \checkmark & 33.46/0.9302 & 40.21/0.9766 
  \\
  (c)          & \checkmark & \checkmark &  & \checkmark & \checkmark & 31.67/0.9051 &  38.06/0.9681 
  \\
  
  (d)          & \checkmark & \checkmark  & \checkmark &  & \checkmark & 33.41/0.9305  & 39.98/0.9758
\\
(e)          & \checkmark & \checkmark  & \checkmark & \checkmark &  & 33.43/0.9311 & 39.91/0.9759
\\
  WTDUN          & \checkmark  & \checkmark  & \checkmark & \checkmark &  \checkmark &   \textbf{33.58}/\textbf{0.9371} & \textbf{40.22}/\textbf{0.9784}
  \\
  \bottomrule
  \end{tabular}
\label{ablation11}
\end{table*}
\begin{table}[!t]
\small
\centering
\caption{The effect of multilevel wavelet transform on reconstruction quality. The sampling ratio of CS is 10\%, 30\%, and 50\% on the Set11 dataset. The best results are highlighted in bold.}
\begin{tabular}{c|c|c|c} 
\toprule
\multirow{2}{*}{METHOD} & \multicolumn{3}{c}{PSNR(dB)/SSIM} \\
\cline{2-4}  
   & {10\%} & {30\%}  & {50\%}\\
\toprule
Ours-level1    & 29.50/0.8791 & 35.26/0.9581 & 40.00/0.9762
\\
Ours-level2    & \textbf{29.53}/\textbf{0.8867} & \textbf{36.27}/\textbf{0.9585} & \textbf{40.22}/\textbf{0.9784}
\\
Ours-level3    & 29.41/0.8808 & 36.18/0.9578 & 40.05/0.9781
\\
  \bottomrule
  \end{tabular}
\label{wdset11}
\end{table}
\begin{table}[!t]
\small
\centering
    \caption{The effect of multilevel wavelet transform on reconstruction quality. The sampling ratio of CS is 10\%, 30\%, and 50\% on the Urban100 dataset. The best results are highlighted in bold.}
\begin{tabular}{c|c|c|c} 
\toprule
\multirow{2}{*}{METHOD} & \multicolumn{3}{c}{PSNR(dB)/SSIM} \\
\cline{2-4}  
   & {10\%} & {30\%}  & {50\%}\\
\toprule
Ours-level1    &  26.52/0.8265 & 31.83/0.9390 & 37.01/0.9702
\\
Ours-level2    & \textbf{26.58}/\textbf{0.8358} & \textbf{33.18}/\textbf{0.9431} & \textbf{37.39}/\textbf{0.9738}
\\
Ours-level3    & 26.42/0.8297 & 33.17/0.9430 & 37.26/0.9732
\\
  \bottomrule
  \end{tabular}
\label{wdurban100}
\end{table}


\begin{table*}[!t]
\small
\centering
\caption{PSNR/SSIM of models with different reconstruction numbers $k$ on Urban100 at the CS ratio of 10\%. The best results are highlighted in bold.}
\begin{tabular}{c|c|c|c|c} %
\toprule
\multirow{2}{*}{METHOD} & \multicolumn{4}{c}{PSNR(dB)/SSIM} \\
\cline{2-5}   
   & {$k$  = 1} & {$k$  = 3}  & {$k$  = 6}& {$k$  = 9}\\
\toprule
ISTA~\cite{ref04}    & 22.03/0.6502 & 23.07/0.7001 & 23.48/0.7176 & 23.56/0.7230
\\
ISTA++~\cite{ref6}    & 17.45/0.3585 & 21.87/0.6322 & 23.29/0.7072 & 24.14/0.7421 \\
AMP-Net~\cite{ref3}    & \textbf{25.03}/\textit{0.7842} & \textit{25.63}/\textit{0.8042} & \textit{25.96}/\textit{0.8133} &\textit{26.05}/\textit{0.8156}
\\
Ours    & \textit{24.85}/\textbf{0.7897} & \textbf{26.03}/\textbf{0.8230} & \textbf{26.36}/\textbf{0.8309} & \textbf{26.58}/\textbf{0.8358}
\\
  \bottomrule
  \end{tabular}
\label{k_100}
\end{table*}  
\begin{table}[!t]
\small
\centering
\caption{Model complexity comparison between WTDUN and other CS methods. PN and PM are the number of the learnable matrix parameters and total parameters, respectively. Running time is computed at 50\% CS ratio.}
\begin{tabular}{c|c|c|c} 
\toprule
\multirow{1}{*}{METHODS} & \multicolumn{3}{c}{Parameters} \\
\cline{2-4}  
   & { PN(Mb)}& {PM(Mb)}&  {Time(s)} \\
\toprule
   ISTA-Net \cite{ref04}  & 1.05  & 2.57 & 0.083 \\ 
  
  OPINE-Net \cite{ref2}   & 2.13  & 4.18 & 0.099  \\
  
  AMP-Net \cite{ref3}     & 2.13  & 5.40 & 0.072 \\
  
  
  COAST \cite{ref5}       & - & 8.56  & 0.093  \\ 
  
  ISTA-Net++ \cite{ref6}  & 2.13  & 5.80 & 0.082\\ 
  
  MADUN \cite{ref7}       & 2.13  & 23.04 & 0.177 \\ 
  
  DPUNet \cite{ref11}     & - & 12.1 & 0.071 \\ 
  Ours                         & 2.50 & 19.72 & 0.170 \\
  \bottomrule
  \end{tabular}
\label{fuzadu}
\end{table}

\section{Experimental Results and Performance Evaluation}   
In this section, our proposed method is first compared with other state-of-the-art methods in terms of both objective reconstruction quality and subjective visual quality. Then a series of ablation experiments are implemented to further evaluate the effects of each functional module in our method. Finally, the model complexity is compared with other methods.
                 
\subsection{Experiment Settings}                                                                
We use $400$ images from the BSD500 dataset for training and validation. In the training process, two training sets are generated for models. (1) Training set with the size $64\times64$ are randomly extracted from images in BSDS500. (2) Training set with the size $128\times128$ are randomly extracted from images in BSDS500. We also unfold the whole testing image in this way during the testing process. We use the PyTorch toolbox and train our model using the Adam solver on an NVIDIA RTX 3090 GPU. All models are trained for $200$ epochs with batch size $32$ and learning rate $0.0001$. Before training, the control parameter $\alpha$ is initialized as 1 and other trainable parameters are initialized randomly.

For testing results, we conduct extensive experiments on some widely used datasets: Set5\cite{ref12}, Set11\cite{ref13}, Set14, Urban100\cite{ref7}. To ensure fairness, we evaluate the performance with two quality evaluation metrics: Peak Signal-to-Noise Ratio(PSNR) and Structural Similarity Index(SSIM). 

\subsection{Comparison with Other Methods}
In this section, we select several typical deep learning-based CS methods for comparison with our method, including DPA-Net\cite{ref1}, BCS-Net\cite{ref4}, OPINE-Net\cite{ref2}, AMP-Net\cite{ref3}, COAST-Net\cite{ref5}, MADUN-Net\cite{ref7}, ISTA-Net++\cite{ref6}, FHDUN-Net\cite{ref9}, ULAMP-Net\cite{ref10}, DPUNet\cite{ref11}, SODAS-Net\cite{ref42} and DPC-DUN~\cite{ref46}. Apart from some experimental results provided by the authors, the results of the other comparison methods are retrained on the Ubuntu 16.04 system under the PyTorch 1.7 framework and use Python 3.7. Moreover, we have also compared our method with some model-based methods on the Set11 dataset, as shown in Table.~\ref{ms_sample}. The source code of comparison methods comes from the official code published by their authors. \Cref{tb1,tb2,tb3,tb4} show the test results of WTDUN and other methods on Set5, Set11, Set14, and Urban100. These tables contain the average PSNR(dB)/SSIM where the best is marked in bold and show the reconstruction results at different sampling rates of \{50\%, 40\%, 30\%, 20\%, 10\%\}.
\subsubsection{Reconstruction Quality}
From \Cref{tb1,tb2,tb3,tb4}, it can be seen that WTDUN can achieve higher PSNR than those deep learning-based CS methods at all sampling ratios. As for SSIM, our WTDUN is better than most methods. It can be seen that MADUN-Net and ISTA-Net++ perform worse at low CS ratios of 10\%. DPA-Net does not work well at all CS ratios. Compared to other methods, for example, the gain of our method with PSNR is about 1.01$ \sim $5.57 dB at 50\% CS ratio and 0.20$ \sim $2.22 dB at 10\% CS ratio on dataset Urban100. The gain of our method with SSIM is about 0.0021$ \sim $0.0301 at 50\% CS ratio and 0.0057$ \sim $0.0561 at 10\% CS ratio on Urban100.
\subsubsection{Visual Effect}
The visual quality comparison of different CS methods on test images is shown in Fig.~\ref{set5_img}, Fig.~\ref{set11_img}, Fig.~\ref{set14_img} and Fig.~\ref{urban100_img} at 10\% CS ratio. It can be seen that the reconstructed images from COAST and ISTA-Net++ have obvious blocking artifacts. This is because these methods only focus on the individual reconstruction of each block and do not consider the correlations between neighboring image blocks. On the contrary, there are no obvious blocking artifacts in the reconstructed images generated by MADUN, OPINE-Net, AMP-Net, and our method, since these models all perform denoising and deblocking operations on the full image at each reconstruction stage. Compared with COAST, ISTA-Net++, MADUN, OPINE-Net, and AMP-Net algorithms, images recovered by our WTDUN have richer texture details and sharper edges than other methods as shown in Fig.~\ref{set11_img} and Fig.~\ref{urban100_img}. Therefore, our WTDUN has a stronger ability to reconstruct high-quality images compared to other state-of-the-art methods.
\subsection{Ablation Study}   
To evaluate the contribution of each component in our WTDUN, we design several variants of the proposed model in which some functional modules are selectively discarded or replaced. Table.~\ref{ablation11} and Table.~\ref{ablation100} show comparative experimental results on Set11 and Urban100 at 20\% and 50\% CS ratios, which include 5 functional modules. Table.~\ref{wdset11} and Table.~\ref{wdurban100} show the comparison of performing wavelet decomposition at different levels. Table.~\ref{k_100} demonstrates the effect of different reconstruction phase numbers $k$ on Urban100 and Set11 at 10\% CS ratio. These \Cref{tb1,tb2,tb3,tb4} show the effect of different patch sizes of image blocks on reconstruction quality on these datasets: Set5, Set11, Set14, and Urban100.
\subsubsection{Validating the Capability of Wavelet Domain Adaptive Sampling}
In this subsection, we validate the capability of the wavelet domain adaptive sampling (WAS) method. CS measurements allocation method is used to adaptively allocate CS measurements according to the target sampling ratio. To test the effect of WAS, we train our network in settings with and without WAS on Urban100 and Set11, as shown in Table.~\ref{ablation100} and Table.~\ref{ablation11}. Compared to the case without WAS, the average PSNR scores of WTDUN can be improved by about $1.5 \sim 2.5$ dB at $20\%$ and $50\%$ CS ratios. The average SSIM scores of our WTDUN can be improved by about $0.01 \sim 0.05$ at $20\%$ and $50\%$ CS ratios with WAS. Ablation results demonstrate that WAS can effectively improve image reconstruction quality.

\subsubsection{Validating the Capability of Wavelet Tree Structure Prior}
In this subsection, we validate the capability of the wavelet tree-structured reconstruction model. The wavelet tree-structured prior (WTP) is used to fully exploit the correlation within multi-scale subbands. To test the effect of WTP, we trained our model in settings with and without WTP. As shown in Table.~\ref{ablation100} and Table.~\ref{ablation11}, the PSNR scores of our WTDUN with WTP are improved by about 0.08 $\sim$ 0.2 dB. It can be seen that our model can obtain higher PSNR/SSIM when the wavelet tree-structured prior is integrated into our framework. 
\subsubsection{Validating the Capability of Deblocking, MC, and CAC}
In this subsection, we validate the capability of the Deblock module, memory context (MC) module and cross-attention (CAC) module. The comparison results are presented in Table.~\ref{ablation100} and Table.~\ref{ablation11}. Of the three mentioned modules, the deblock module brings the largest improvement in PSNR/SSIM, with a gain of about 1 $ \sim $ 2.5 dB for PSNR and 0.02 $ \sim $ 0.1 for SSIM. The gain for MC and CAC is almost the same, about 0.2 dB for PSNR and 0.01 for SSIM. 
\subsubsection{Validating the Influence Image Patch Size}
In this subsection, we validate the influence of different image patch sizes. These \Cref{tb1,tb2,tb3,tb4} show that the reconstruction performance can be improved with the increase of the block size. {Larger patch size can help our model capture more contextual information, which is vital for reconstruction methods. The contextual information enables the method to more effectively estimate and fill in missing or corrupted areas of the image, resulting in a more accurate and visually coherent reconstruction.
\subsubsection{Validating the Capability of Wavelet Decomposition at Different Levels}
In this subsection, we validate the capability of wavelet decomposition at different levels on Set11 and Urban100. Table.~\ref{wdurban100} and Table.~\ref{wdset11} show that the best reconstruction results can be obtained at two-level wavelet decomposition. Multi-level wavelet decomposition can increase image sparsity, which can lead to better reconstruction quality. Therefore, the PSNR/SSIM of the two-level and three-level wavelet decomposition are better than the one-level wavelet decomposition. Since we use $64 \times 64$ image blocks as input to our model, we can get some $8 \times 8$ wavelet subbands after three-level decomposition. Because the wavelet subbands of size $8 \times 8$ are too small for deep networks, the useful feature cannot be effectively extracted, resulting in poor reconstruction quality. Thus, the PSNR/SSIM of the two-level wavelet decomposition is better than that of the three-level wavelet decomposition. 
\begin{figure*}
    \centering
    \includegraphics[width=0.45\linewidth]{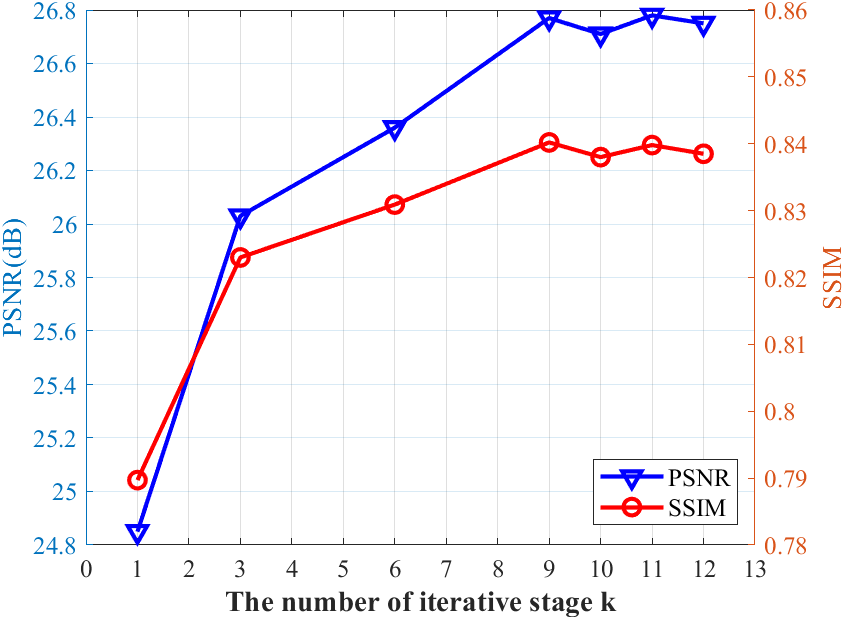}
    \caption{The PSNR/SSIM curve with the increase of iterative stage $K$.}
    \label{stage_k}
\end{figure*}
\subsubsection{Validating the Capability of Different Reconstruction Stage $K$}
To investigate the selection of $K=9$ as the optimal value for our CS model, we conduct a series of comparative experiments. These experiments show that when $K$ goes beyond $9$, the improvement in PSNR/SSIM becomes marginal. To balance the performance and complexity, $K = 9$ is selected as the number of iterations in our CS reconstruction model, as depicted in Fig.~\ref{stage_k} and Table.~\ref{k_100}.

\subsection{Model Complexity}
In this section, we compare the model complexity of different CS approaches in terms of parameter capacity and time complexity. The average running time is used to evaluate the actual reconstruction efficiency. Note that the average running time and the number of parameters are tested by reconstructing 9 test images of size $256 \times 256$. 
\subsubsection{Parameter Capacity}
Table~\ref{fuzadu} shows the number of sampling matrix parameters for ISTA-Net, ISTA-Net++, OPINE-Net, AMP-Net, COAST, MADUN, DPUNet and our method. For a normal task, our method can achieve better reconstruction quality and faster reconstruction speed compared to MADUN with a similar number of parameters. Even though our method samples with larger patches ($64 \times 64$), the number of parameters we use only increases by about 0.4Mb parameters rather than nearly 4 times parameters compared to small-size patches ($33 \times 33$). 

\subsubsection{Time Complexity}
As shown in Table~\ref{fuzadu}, the average running time of deep learning methods on GPU is less than 0.2 seconds. The reconstruction process of traditional CS methods is implemented iteratively until convergence or the maximum iteration step is reached. As a result, the reconstruction speed of traditional methods is lower than deep learning methods and deep unfolding methods. Moreover, GPUs have advantages over CPUs in terms of computational power and parallel computing capability. This can greatly reduce the running time of deep learning-based CS and deep unfolding approaches. 

All CS models shown in Table~\ref{fuzadu} have different time complexities. Owing to the superior computing power of GPUs, the slight difference in the running time of these methods is not significant. Since there is only a small difference in the running time of models, image reconstruction quality is more important for deep learning-based methods. In summary, our method achieves a better accuracy-complexity trade-off than other state-of-the-art methods.

\section{Conclusions}
We propose a novel wavelet domain framework named WTDUN for image CS, which enables simultaneous sampling based on subband characteristics and reconstruction guided by wavelet tree-structured prior. To achieve better CS sampling, we have designed an algorithm that allocates CS measurements based on the importance of each subband. Additionally, we have developed a wavelet domain sampling method to achieve adaptive sampling, thereby effectively enhancing the acquisition capability of sampling information. For CS reconstruction, we propose a tree-structured prior guided unfolding network. This innovative method effectively maintains a similar structure among wavelet subbands within a group by fully exploiting their inherent sparsity. By leveraging the advantages of tree-structured sparsity, our method significantly enhances the quality of image reconstruction compared to other state-of-the-art CS methods. Furthermore, as we increase the number of reconstruction modules, we observe further improvements in the reconstructed results. Our future work will focus on extending our model to exploit structural differences between blocks and facilitate effective interaction among different stages of reconstruction.
\vspace{-0.2cm}

\begin{acks}
This work was partially supported by the National Key R$\&$D Program of China (No. 2021ZD0111902), the National Natural Science Foundation of China (62272016, 62372018, U21B2038), and the National Science Foundation through grant DMS-2304489. 
\end{acks}


\bibliographystyle{ACM-Reference-Format}
\bibliography{sample-acmsmall}



\end{document}